\documentclass[conference]{IEEEtran}
\IEEEoverridecommandlockouts

\usepackage[
  paper=letterpaper,
  top=0.95in,
  bottom=1.0in,
  left=0.75in,
  right=0.75in,
  headheight=0pt,
  headsep=0pt,
  footskip=0.5in
]{geometry}

\usepackage{amsmath,amssymb,amsfonts}   %
\usepackage{textcomp}                   %

\usepackage{graphicx}                   %
\usepackage[dvipsnames,svgnames,table]{xcolor}                     %

\usepackage{array}                      %
\newcolumntype{H}{>{\setbox0=\hbox\bgroup}c<{\egroup}@{}} %
\usepackage{booktabs}                   %
\usepackage{siunitx}                    %
  \DeclareSIUnit{\dBm}{dBm}
  \DeclareSIUnit{\dBi}{dBi}
  \DeclareSIUnit{\dBsm}{dBsm}

\usepackage{soul}                       %

\usepackage[backend=biber,
            language=english,
            autolang=other,
            style=ieee]{biblatex}       %
\usepackage[dvipsnames,svgnames,table]{xcolor} %
\usepackage[hidelinks]{hyperref}               %
\usepackage[capitalize]{cleveref}                          %

\usepackage{algorithmic}                %

\usepackage{pgfplots}                   %
  \pgfplotsset{compat=newest}
\usepgfplotslibrary{groupplots,dateplot} %
\usepackage{pgfplotstable}              %
  \pgfplotsset{compat=1.18}              %

\usetikzlibrary{%
  patterns,          %
  shapes.arrows,     %
  spy,               %
  external,          %
  calc,              %
  shadings,          %
  shadows.blur,      %
  decorations.pathreplacing %
}

\usepackage[xindy,nonumberlist]{glossaries} %
  \makenoidxglossaries            %
  
\newacronym{2d}{2D}{two-dimensional}
\newacronym{3d}{3D}{three-dimensional}
\newacronym{3gpp}{3GPP}{3rd generation partnership project}
\newacronym{3msk}{3MSK}{3-level minimum-shift-keying}
\newacronym{4g}{4G}{fourth-generation}
\newacronym{5g}{5G}{fifth-generation}
\newacronym{6dof}{6DoF}{six degrees of freedom}
\newacronym{6g}{6G}{sixth-generation}
\newacronym{abp}{ABP}{authentication by personalisation}
\newacronym{ac}{AC}{alternating current}
\newacronym{aci}{ACI}{adjacent channel interference}
\newacronym{aclr}{ACLR}{adjacent channel leakage ratio}
\newacronym{acpr}{ACPR}{adjacent channel power ratio}
\newacronym{adc}{ADC}{analog-to-digital converter}
\newacronym{adr}{ADR}{adaptive data rate}
\newacronym{aec}{AEC}{aluminum electrolytic capacitor}
\newacronym{aes}{AES}{Advanced Encryption Standard}
\newacronym{afe}{AFE}{analog front end}
\newacronym{ag}{AG}{array gain}
\newacronym{agc}{AGC}{automatic gain controller}
\newacronym{agps}{A-GPS}{Assisted Global Positioning System} %
\newacronym{ai}{AI}{artificial intelligence}
\newacronym{alk}{ALK}{Alkaline}
\newacronym{am}{AM}{amplitude modulation}
\newacronym{amam}{AM/AM}{amplitude modulation to amplitude modulation}
\newacronym{amc}{AMC}{automatic modulation classification}
\newacronym{amp}{AMP}{approximate message passing}
\newacronym{ampm}{AM/PM}{amplitude modulation to phase modulation}
\newacronym{aoa}{AOA}{angle-of-arrival}
\newacronym{aod}{AOD}{angle-of-departure}
\newacronym{ap}{AP}{access point}
\newacronym{api}{API}{application program interface}
\newacronym{apt}{APT}{acoustic power transfer}
\newacronym{apu}{APU}{access point unit}
\newacronym{ar}{AR}{augmented reality}
\newacronym{arima}{ARIMA}{Auto-Regressive Integrated Moving Average}
\newacronym{arp}{ARP}{Antenna Reference Point}
\newacronym{asic}{ASIC}{application specific integrated circuit}
\newacronym{ask}{ASK}{amplitude-shift keying}
\newacronym{at}{AT}{ATtention}
\newacronym{auv}{AUV}{autonomous underwater vehicle}
\newacronym{awg}{AWG}{arbitrary waveform generator}
\newacronym{awgn}{AWGN}{additive white Gaussian noise}
\newacronym{baw}{BAW}{bulk acoustic wave}
\newacronym{bb}{BB}{base-band}
\newacronym{bc}{BC}{backscatter communication}
\newacronym{bcjr}{BCJR}{Bahl-Cocke-Jelinek-Raviv}
\newacronym{bd}{BD}{backscatter device}
\newacronym{be}{BE}{Belgium}
\newacronym{ber}{BER}{bit error rate}
\newacronym{bf}{BF}{beamforming}
\newacronym{bga}{BGA}{ball grid array}
\newacronym{bldc}{BLDC}{brushless DC}
\newacronym{bler}{BLER}{block error rate}
\newacronym{bod}{BOD}{Brown-Out Detection}
\newacronym{bom}{BOM}{bill of materials}
\newacronym{bpsk}{BPSK}{binary phase-shift keying}
\newacronym{brp}{BRP}{beam refinement process}
\newacronym{bs}{BS}{base station}
\newacronym{bu}{BU}{booster unit}
\newacronym{bw}{BW}{bandwidth}
\newacronym{ca}{CA}{carrier emitter}
\newacronym{cad}{CAD}{channel activity detection}
\newacronym{cars}{CARS}{calibration reference signal}
\newacronym{cbm}{CBM}{condition based maintenance}
\newacronym{cc}{CC}{constant current}
\newacronym{ccdf}{CCDF}{complementary cumulative distribution function}
\newacronym{ccnn}{CCNN}{circular convolutional neural network}
\newacronym{ccs}{CCS}{correlative channel sounder}
\newacronym{cdf}{CDF}{cumulative distribution function}
\newacronym{cdma}{CDMA}{code division-multiple access}
\newacronym{cdrx}{CDRX}{connected mode DRX}
\newacronym{ce}{CE}{coverage enhancement}
\newacronym{ced}{CED}{cumulative energy density}
\newacronym{ceofdm}{CE-OFDM}{constant-envelope OFDM}
\newacronym{cf}{CF}{cell-free}
\newacronym{cfmmimo}{CF-mMIMO}{cell-free massive MIMO}
\newacronym{cfo}{CFO}{carrier frequency offset}
\newacronym{cir}{CIR}{channel impulse response}
\newacronym{cla}{CLA}{closed-loop approach}
\newacronym{clk}{CLK}{clock}
\newacronym{cmos}{CMOS}{complementary metal oxide semiconductor}
\newacronym{cnn}{CNN}{convolutional neural network}
\newacronym{co}{CO}{combinatorial optimization}
\newacronym{cordic}{CORDIC}{coordinate rotation digital computer}
\newacronym{cost}{COST}{commercial off-the-shelf}
\newacronym{cots}{COTS}{commercial off-the-shelf}
\newacronym{cp}{CP}{cyclic prefix}
\newacronym{cpe}{CPE}{common phase error}
\newacronym{cpfsk}{CPFSK}{continuous phase frequency shift keying}
\newacronym{cpm}{CPM}{continuous phase modulation}
\newacronym{cpt}{CPT}{capacitive power transfer}
\newacronym{cpu}{CPU}{central-processing unit}
\newacronym{cpw}{CPW}{coplanar waveguide}
\newacronym{cqi}{CQI}{channel quality indicator}
\newacronym{cr}{CR}{coding rate}
\newacronym{crc}{CRC}{cyclic redundancy check}
\newacronym{crlb}{CRLB}{Cram\'er-Rao lower bound}
\newacronym{crs}{CRS}{cell reference signal}
\newacronym{cs}{CS}{compressed sensing}
\newacronym{cse}{CSE}{channel state estimation}
\newacronym{csi}{CSI}{channel state information}
\newacronym{csp}{CSP}{contact service point}
\newacronym{css}{CSS}{chirp spread spectrum}
\newacronym{cu}{CU}{central unit}
\newacronym{cv}{CV}{constant voltage}
\newacronym{cw}{CW}{continuous wave}
\newacronym{d2d}{D2D}{device-to-device}
\newacronym{dab}{DAB}{Digital Audio Broadcasting}
\newacronym{dac}{DAC}{digital-to-analog converter}
\newacronym{daq}{DAQ}{data acquisition system}
\newacronym{das}{DAS}{distributed antenna systems}
\newacronym{dbpsk}{DBPSK}{Differential Binary Phase Shift Keying}
\newacronym{dc}{DC}{direct current}
\newacronym{dcc}{DCC}{dynamic cooperation clustering}
\newacronym{ddc}{DDC}{digital down conversion}
\newacronym{de}{DE}{drain efficiency}
\newacronym{dft}{DFT}{discrete Fourier transform}
\newacronym{dftsofdm}{DFT-s-OFDM}{discrete Fourier Transform spread OFDM}
\newacronym{dftsofdmfdss}{DFT-s-OFDM-FDSS}{DFT-s-OFDM with frequency-domain spectral shaping}
\newacronym{dftsofdmfdssse}{DFT-s-OFDM-FDSS-SE}{DFT-s-OFDM with FDSS and spectral extension}
\newacronym{dl}{DL}{downlink}
\newacronym{dlc}{DLC}{Distributed Laser Charging}
\newacronym{dli}{DLI}{direct link interference}
\newacronym{dlt}{DLT}{Distributed Ledger Technology}
\newacronym{dm}{DM}{diffuse multipath}
\newacronym{dma}{DMA}{Direct Memory Access}
\newacronym{dmac}{DMAC}{Direct Memory Access Controller}
\newacronym{dmc}{DMC}{diffuse multipath component}
\newacronym{dmimo}{D-MIMO}{distributed MIMO}
\newacronym{dnn}{DNN}{deep neural network}
\newacronym{doa}{DOA}{direction-of-arrival}
\newacronym{dp}{DP}{differencial privacy}
\newacronym{dpd}{DPD}{digital pre-distortion}
\newacronym{dpdk}{DPDK}{Data Plane Development Kit}
\newacronym{dram}{DRAM}{dynamic random-access memory}
\newacronym{drcs}{$\Delta$RCS}{differential-radar cross section}
\newacronym{drx}{DRX}{Discontinuous Reception Mode}
\newacronym{dsb}{DSB}{double-sideband}
\newacronym{dsl}{DSL}{digital subscriber line}
\newacronym{dsp}{DSP}{digital signal processing}
\newacronym{dss}{DSS}{Dataset Storage Standard}
\newacronym{duc}{DUC}{digital up-converter}
\newacronym{dvb}{DVB}{Digital Video Broadcasting}
\newacronym{e2e}{E2E}{end-to-end}
\newacronym{easa}{EASA}{European Union Aviation Safety Agency}
\newacronym{ebg}{EBG}{electromagnetic bandgap}
\newacronym{ebw}{EBW}{excess bandwidth}
\newacronym{ec}{EC}{European Commission}
\newacronym{ecc}{ECC}{elliptic curve cryptography}
\newacronym{ecdf}{eCDF}{empirical cumulative distribution function}
\newacronym{ecdlp}{ECDLP}{elliptic curve discrete logarithm problem}
\newacronym{ecsp}{ECSP}{edge computing service point}
\newacronym{edlc}{EDLC}{electrostatic double-layer capacitors}
\newacronym{edrx}{eDRX}{Extended Discontinuous Reception Mode}
\newacronym{ee}{EE}{energy efficiency}
\newacronym{egprs}{EGPRS}{Enhanced Data Rates for GSM Evolution}
\newacronym{eh}{EH}{energy harvesting}
\newacronym{eirp}{EIRP}{equivalent isotropically radiated power}
\newacronym{em}{EM}{electromagnetic}
\newacronym{embb}{eMBB}{enhanced Mobile Broadband}
\newacronym{en}{EN}{energy neutral}
\newacronym{end}{END}{energy-neutral device}
\newacronym{enob}{ENOB}{effective number of bits}
\newacronym{eol}{EoL}{end of life}
\newacronym{ep}{EP}{energy profiler}
\newacronym{epd}{EPD}{electronic paper display}
\newacronym{epu}{EPU}{edge processing unit}
\newacronym{er}{ER}{Energy Receiver}
\newacronym{erp}{ERP}{effective radiation power}
\newacronym[plural=ESCs,firstplural=electronic speed controllers (ESCs)]{esc}{ESC}{electronic speed control}
\newacronym{esd}{ESD}{electrostatic discharge}
\newacronym{esl}{ESL}{electronic shelf label}
\newacronym{esr}{ESR}{equivalent series resistance}
\newacronym{et}{ET}{Energy Transmitter}
\newacronym{etsi}{ETSI}{European Telecommunications Standards Institute}
\newacronym{evd}{EVD}{eigenvalue decomposition}
\newacronym{evm}{EVM}{Error Vector Magnitude}
\newacronym{ewlb}{eWLB}{Embedded Wafer Level Ball Grid Array}
\newacronym{fa}{FA}{federation anchor}
\newacronym{fair}{FAIR}{Findability, Accessibility, Interoperability, and Reuse of digital assets}
\newacronym{fcf}{FCF}{frequency correlation function}
\newacronym{fd}{FD}{front-haul distance}
\newacronym{fdd}{FDD}{frequency-division duplexing}
\newacronym{fde}{FDE}{frequency domain equalizer}
\newacronym{fdm}{FDM}{frequency-division multiplexing}
\newacronym{fdma}{FDMA}{frequency division multiple access}
\newacronym{fdss}{FDSS}{frequency-domain spectral shaping}
\newacronym{fem}{FEM}{finite element analysis}
\newacronym{fembb}{feMBB}{further enhanced mobile broadband}
\newacronym{fft}{FFT}{fast Fourier transform}
\newacronym{fh}{FH}{fronthaul}
\newacronym{fhss}{FHSS}{frequency hopping spread spectrum}
\newacronym{fifo}{FIFO}{First In, First Out}
\newacronym{fim}{FIM}{Fisher information matrix}
\newacronym{fits}{FITS}{Flexible Image Transport System}
\newacronym{fl}{FL}{federated learning}
\newacronym{fom}{FoM}{figure of merit}
\newacronym{fov}{FOV}{field of view}
\newacronym{fpga}{FPGA}{field-programmable gate array}
\newacronym{fr1}{FR1}{frequency range 1}
\newacronym{fr2}{FR2}{frequency range 2}
\newacronym{fram}{FRAM}{Ferroelectric Random Access Memory}
\newacronym{fsk}{FSK}{frequency shift keying}
\newacronym{fspl}{FSPL}{free space path loss}
\newacronym{fss}{FSS}{frequency selective surface}
\newacronym{gb}{GB}{grant-based}
\newacronym{gdpr}{GDPR}{general data protection regulation}
\newacronym{gf}{GF}{grant-free}
\newacronym{gmsk}{GMSK}{Gaussian minimum-shift keying}
\newacronym{gnb}{gNB}{Next Generation Node B}
\newacronym{gni}{GNI}{gross national income}
\newacronym{gnn}{GNN}{graph neural network}
\newacronym{gnss}{GNSS}{global navigation satellite system}
\newacronym{gpclk}{GPCLK}{general purpose clock}
\newacronym{gpio}{GPIO}{General-Purpose Input/Output}
\newacronym{gpl}{GPL}{GNU General Public License}
\newacronym{gprs}{GPRS}{General Packet Radio Services}
\newacronym{gps}{GPS}{Global Positioning System}
\newacronym{gpu}{GPU}{graphical processing unit}
\newacronym{grc}{GRC}{GNU Radio Companion}
\newacronym{gscm}{GSCM}{geometry‐based stochastic model}
\newacronym{gsm}{GSM}{Global System for Mobile Communications}  %
\newacronym{gspm}{GSpM}{Generalized spatial modulation}
\newacronym{gwp}{GWP}{Global Warming Potential}
\newacronym{harq}{HARQ}{hybrid automatic repeat request}
\newacronym{hat}{HAT}{hardware attached on top}
\newacronym{hcs}{HCS}{human-centric services}
\newacronym{hdf5}{HDF5}{Hierarchical Data Format version 5}
\newacronym{hfss}{HFSS}{High Frequency Simulator Software}
\newacronym{hmd}{HMD}{head-mounted display}
\newacronym{hpbm}{HPBM}{half power beam width}
\newacronym{HyMPRo}{HyMPRo}{Hybrid Multi-Path Routing algorithm}
\newacronym{i.i.d.}{i.i.d.}{independent and identically distributed}
\newacronym{i2c}{I2C}{Inter-Integrated Circuit}
\newacronym{iaq}{IAQ}{Indoor Air Quality}
\newacronym{ib}{IB}{in-band}
\newacronym{ibbc}{IBBC}{inter-band beam configuration}
\newacronym{ibo}{IBO}{input back-off}
\newacronym{ic}{IC}{integrated circuit}
\newacronym{iccs}{ICCS}{Ilmsens correlative channel sounder}
\newacronym{ici}{ICI}{intercarrier interference}
\newacronym{icnirp}{ICNIRP}{International Commission on Non-Ionizing Radiation Protection}
\newacronym{id}{ID}{information decoding}
\newacronym{idft}{IDFT}{inverse discrete Fourier transform}
\newacronym{idxm}{IDXM}{index modulation}
\newacronym{if}{IF}{intermediate-frequency}
\newacronym{ifft}{IFFT}{inverse fast-Fourier-transform}
\newacronym{iid}{i.i.d.}{independently and identically distributed}
\newacronym{iis}{IIS}{integrated information system}
\newacronym{im}{IM}{intermodulation}
\newacronym{imd}{IMD}{intermodulation distortion}
\newacronym{imu}{IMU}{inertial measurement unit}
\newacronym{inh}{InH}{indoor hotspot office}
\newacronym{io}{IO}{input/output}
\newacronym{ioe}{IoE}{Internet of Everything}
\newacronym{iot}{IoT}{Internet of Things}
\newacronym{ipt}{IPT}{inductive power transfer}
\newacronym{ipy}{IPY}{Interventions per Year}
\newacronym{iq}{IQ}{in-phase and quadrature}
\newacronym{iqi}{IQI}{IQ imbalance}
\newacronym{ir}{IR}{infrared}
\newacronym{isi}{ISI}{intersymbol interference}
\newacronym{ism}{ISM}{industrial, scientific and medical}
\newacronym{isp}{ISP}{internet service provider}
\newacronym{itu}{ITU}{International Telecommunication Union}
\newacronym{jesd}{JESD}{Joint Electron Devices Engineering Council}
\newacronym{jfet}{JFET}{junction field effect transistor}
\newacronym{kpi}{KPI}{key performance indicator}
\newacronym{ktofdm}{KT-DFT-s-OFDM}{known-tail-DFT-s-OFDM}
\newacronym{kvi}{KVI}{key value indicator}
\newacronym{larva}{LARVA}{LARge Virtual Array}
\newacronym{lca}{LCA}{life cycle assessment}
\newacronym{lco}{LCO}{lithium cobalt oxide}
\newacronym{ldo}{LDO}{Low-dropout voltage regulator}
\newacronym{ldpc}{LDPC}{low-density parity-check}
\newacronym{led}{LED}{Light Emitting Diode}
\newacronym{less}{LESS}{Low Energy Scheduler Solution}
\newacronym{lfp}{LFP}{lithium iron phosphate}
\newacronym{lib}{LIB}{Lithium-Ion Battery}
\newacronym{lic}{LIC}{lithium-ion capacitor}
\newacronym{lid}{LID}{Lithium Iron Disulfide}
\newacronym{lidar}{LiDAR}{light detection and ranging}
\newacronym{liion}{Li-ion}{lithium-ion}
\newacronym{lipo}{LiPo}{lithium polymer}
\newacronym{lis}{LIS}{large intelligent surface}
\newacronym{llh}{LLH}{log-likelihood}
\newacronym{lls}{LLS}{link-level simulation}
\newacronym{lmd}{LMD}{Lithium Manganese Dioxide}
\newacronym{lmmse}{LMMSE}{least minimum mean square error}
\newacronym{lmo}{LMO}{lithium ion manganese oxide}
\newacronym{lna}{LNA}{low-noise amplifier}
\newacronym{lo}{LO}{local oscillator}
\newacronym{lora}{LoRa}{long range}
\newacronym{lorawan}{LoRaWAN}{long-range wide-area network}
\newacronym{los}{LoS}{line-of-sight}
\newacronym{lp}{LP}{linear programming}
\newacronym{lpf}{LPF}{low-pass filter}
\newacronym{lpt}{LPT}{laser power transfer}
\newacronym{lpwa}{LPWA}{Low Power Wide Area}
\newacronym{lpwan}{LPWAN}{low-power wide-area network}
\newacronym{lpwans}{LPWANs}{Low-Power Wide-Area Networks}
\newacronym{lqi}{LQI}{link quality indicator}
\newacronym{lrelu}{LReLU}{leaky rectified linear unit}
\newacronym{lrt}{LRT}{likelihood-ratio test}
\newacronym{ls}{LS}{least squares}
\newacronym{lsa}{LSA}{large synthetic array}
\newacronym{lsf}{LSF}{large-scale fading}
\newacronym{lsfc}{LSFC}{large-scale fading component}
\newacronym{lstm}{LSTM}{Long Short-Term Memory}
\newacronym{ltc}{LTC}{lithium thionyl chloride}
\newacronym{lte}{LTE}{Long Term Evolution}
\newacronym{lti}{LTI}{linear time-invariant}
\newacronym{lto}{LTO}{lithium titanate}
\newacronym{lusta}{LUSTA 5G }{Logistique mUltimodale Sécuritaire Téléopérée \& Autonome 5G}
\newacronym{m2m}{M2M}{machine to machine}
\newacronym{mac}{MAC}{Medium Access Control}
\newacronym{mate}{MATE}{millimeter-wave MIMO testbed}
\newacronym{mc}{MC}{Monte Carlo}
\newacronym{mcl}{MCL}{Maximum Coupling Loss}
\newacronym{mcs}{MCS}{modulation and coding scheme}
\newacronym{mcu}{MCU}{microcontroller unit}
\newacronym{mec}{MEC}{multi-access edge computing}
\newacronym{mems}{MEMS}{micro-electromechanical systems}
\newacronym{mf}{MF}{matched filter}
\newacronym{mimo}{MIMO}{multiple-input multiple-output}
\newacronym{miso}{MISO}{multiple-input single-output}
\newacronym{ml}{ML}{machine learning}
\newacronym{mlp}{MLP}{multilayer perceptron}
\newacronym{mmic}{MMIC}{monolithic microwave integrated circuit}
\newacronym{mmimo}{mMIMO}{massive MIMO}
\newacronym{mmse}{MMSE}{minimum mean square error}
\newacronym{mmtc}{mMTC}{massive machine-typed communication}
\newacronym{mmwave}{mmWave}{millimeter wave}
\newacronym{mn}{MN}{matching network}
\newacronym{mosfet}{MOSFET}{metal-oxide semiconductor field effect transistor}
\newacronym{mpc}{MPC}{multipath component}
\newacronym{mppt}{MPPT}{maximum power point tracking}
\newacronym{mr}{MR}{maximum ratio}
\newacronym{mrc}{MRC}{maximum ratio combining}
\newacronym{mrc_em}{MRC}{maximum ratio combining}
\newacronym{mrc_EM}{MRC}{Magnetic Resonance Coupling}
\newacronym{mrt}{MRT}{maximum ratio transmission}
\newacronym{mse}{MSE}{mean square error}
\newacronym{msk}{MSK}{Minimum-Shift Keying}
\newacronym{mtc}{MTC}{Machine-Type Communication}
\newacronym{multi-rat}{Multi-RAT}{multiple radio access technology}
\newacronym{multirat}{Multi-RAT}{Multiple Radio Access Technology}
\newacronym{music}{MUSIC}{MUltiple SIgnal Classification}
\newacronym{navauwall}{NAVAUWALL}{AUtomated NAVigation in WALLonia}
\newacronym{nb}{NB}{narrowband}
\newacronym{nbiot}{NB-IoT}{narrowband IoT}
\newacronym{nca}{NCA}{nickel cobalt aluminum}
\newacronym{netcdf}{NetCDF}{Network Common Data Form}
\newacronym{nf}{NF}{noise figure}
\newacronym{nfc}{NFC}{near-field communication}
\newacronym{nfv}{NFV}{network function virtualization}
\newacronym{ngmn}{NGMN}{Next Generation Mobile Networks }
\newacronym{ni}{NI}{National Instruments}
\newacronym{nicd}{NiCd}{nikkel cadmium}
\newacronym{nimh}{NiMH}{nikkel metal hydride}
\newacronym{nlos}{NLoS}{non-line-of-sight}
\newacronym{nmc}{NMC}{nickel manganese cobalt}
\newacronym{nmos}{nMOS}{n-channel metal-oxide semiconductor}
\newacronym{nn}{NN}{neural network}
\newacronym{nnls}{NNLS}{non-negative least squares}
\newacronym{noma}{NOMA}{non-orthogonal multiple access}
\newacronym{np}{NP}{Neyman-Pearson}
\newacronym{npbch}{NPBCH}{Narrowband Physical Broadcast Channel}
\newacronym{npss}{NPSS}{Narrow Band Primay Synchronization Signal}
\newacronym{nr}{NR}{New Radio}
\newacronym{nrs}{NRS}{Narrow Band Reference Signal}
\newacronym{nsss}{NSSS}{Narrowband Secondary Synchronization Signal}
\newacronym{ntp}{NTP}{network time protocol}
\newacronym{oai}{OAI}{OpenAirInterface} %
\newacronym{obw}{OBW}{occupied bandwidth}
\newacronym{ofdm}{OFDM}{orthogonal frequency-division multiplexing}
\newacronym{ofdma}{OFDMA}{orthogonal frequency-division multiple access}
\newacronym{ofdmim}{OFDM-IM}{OFDM with index modulation}
\newacronym{olos}{OLoS}{obstructed-line-of-sight}
\newacronym{oma}{OMA}{orthogonal multiple access}
\newacronym{oob}{OOB}{out-of-band}
\newacronym{ook}{OOK}{on-off keying}
\newacronym{opbo}{OPBO}{output power backoff}
\newacronym{oran}{O-RAN}{open radio-access network}
\newacronym{os}{OS}{operating system}
\newacronym{ota}{OTA}{over-the-air}
\newacronym{otaa}{OTAA}{over-the-air authentication}
\newacronym{p1}{P1}{Phase 1}
\newacronym{p2}{P2}{Phase 2}
\newacronym{p2p}{P2P}{point-to-point}
\newacronym{pa}{PA}{power amplifier}
\newacronym{pae}{PAE}{power-added efficiency}
\newacronym{pam}{PAM}{pulse amplitude modulation}
\newacronym{pana}{PanA}{Panel A}
\newacronym{panb}{PanB}{Panel B}
\newacronym{papr}{PAPR}{peak-to-average power ratio}
\newacronym{pc}{PC}{pilot count}
\newacronym{pcb}{PCB}{printed circuit board}
\newacronym{pcg}{PCG}{power consumption gain}
\newacronym{pcie}{PCIe}{Peripheral Component Interconnect Express}
\newacronym{pcsi}{PCSI}{perfect channel state information}
\newacronym{pd}{PD}{powered device}
\newacronym{pdcch}{PDCCH}{physical downlink control channel}
\newacronym{pdf}{PDF}{probability density function}
\newacronym{pdp}{PDP}{power delay profile}
\newacronym{pdsch}{PDSCH}{physical downlink shared channel}
\newacronym{pe}{PE}{processing element}
\newacronym{peb}{PEB}{positioning error bound}
\newacronym{per}{PER}{packet error rate}
\newacronym{pet}{PET}{privacy enhancing technology}
\newacronym{pg}{PG}{path gain}
\newacronym{pgd}{PGD}{proximal gradient descent}
\newacronym{phy}{PHY}{physical}
\newacronym{pki}{PKI}{public key infrastucture}
\newacronym{pl}{PL}{path loss}
\newacronym{pla}{PLA}{physically large array}
\newacronym{pll}{PLL}{phase-locked loop}
\newacronym[plural=PMs,firstplural=person months (PMs)]{pm}{PM}{person month}
\newacronym{pmf}{PMF}{polymer microwave fiber}
\newacronym{pmu}{PMU}{power management unit}
\newacronym{pn}{PN}{pseudo-noise}
\newacronym{po}{PO}{phase offset}
\newacronym{poc}{PoC}{proof of concept}
\newacronym{poe}{PoE}{power-over-Ethernet}
\newacronym{pps}{1PPS}{pulse per second}
\newacronym{pr}{PR}{phase reversal}
\newacronym{prb}{PRB}{Physical Resource Block}
\newacronym{prbs}{PRBs}{Physical Resource Blocks}
\newacronym{prs}{PRS}{Peripheral Reflex System}
\newacronym{ps}{PS}{Processing System}
\newacronym{psd}{PSD}{power spectral density}
\newacronym{pse}{PSE}{power sourcing equipment}
\newacronym{psk}{PSK}{phase shift keying}
\newacronym{psm}{PSM}{power saving mode}
\newacronym{pss}{PSS}{primary synchronisation signal}
\newacronym{ptp}{PTP}{precision-time protocol}
\newacronym{ptrs}{PTRS}{Phase-Tracking Reference Signals}
\newacronym{ptw}{PTW}{paging time window}
\newacronym{pv}{PV}{photovoltaic}
\newacronym{pw}{PW}{planar wavefront}
\newacronym{pwm}{PWM}{pulse width modulation}
\newacronym{qam}{QAM}{quadrature amplitude modulation}
\newacronym{qos}{QoS}{quality-of-service}
\newacronym{qpsk}{QPSK}{quadrature phase-shift keying}
\newacronym{qrrls}{QR-RLS}{QR decomposition based recursive least squares}
\newacronym{quadriga}{QuaDRiGa}{QUAsi Deterministic RadIo channel GenerAtor}
\newacronym{ra}{RA}{Random Access}
\newacronym{ram}{RAM}{random-access memory}
\newacronym{ran}{RAN}{radio access network}
\newacronym{rar}{RAR}{Random Access Response}
\newacronym{rat}{RAT}{radio access technology}
\newacronym{rb}{Rb}{Rubidium}
\newacronym{rbs}{RBS}{radio base station}
\newacronym{rbw}{RBW}{resolution bandwidth}
\newacronym{rc}{RC}{raised-cosine}
\newacronym{rcs}{RCS}{radar cross section}
\newacronym{rdl}{RDL}{redistribution layer}
\newacronym{re}{RE}{radio element}
\newacronym{relu}{ReLU}{rectified linear unit}
\newacronym{rf}{RF}{radio frequency}
\newacronym{rfeh}{RFEH}{radio frequency energy harvesting}
\newacronym{rfic}{RFIC}{radio-frequency integrated circuit}
\newacronym{rfid}{RFID}{radio frequency identification}
\newacronym{rfpt}{RFPT}{radio frequency power transfer}
\newacronym{rfsoc}{RFSoC}{Radio Frequency System-on-Chip}
\newacronym{rir}{RIR}{room impulse response}
\newacronym{ris}{RIS}{reflective intelligent surface}%
\newacronym{rllmtc}{RLLMTC}{reliable low latency machine type communication}
\newacronym{rls}{RLS}{recursive least squares}
\newacronym{rms}{RMS}{root-mean-square}
\newacronym{rmse}{RMSE}{root-mean-square error}
\newacronym{rmt}{RMT}{random matrix theory}
\newacronym{rof}{RoF}{radio-over-fiber}
\newacronym{ros}{ROS}{robot operating system}
\newacronym{rpi}{RPi}{Raspberry Pi}
\newacronym{rrc}{RRC}{Radio Resource Connection}
\newacronym{rreq}{RREQ}{route request packet}
\newacronym{rsrp}{RSRP}{Reference Signals Received Power}
\newacronym{rsrq}{RSRQ}{Reference Signal Received Quality}
\newacronym{rss}{RSS}{received signal strength}
\newacronym{rssi}{RSSI}{received signal strength indicator}
\newacronym{rtc}{RTC}{real time clock}
\newacronym{rtf}{RTF}{reader talks first}
\newacronym{rtk}{RTK}{real time kinematics}
\newacronym{rts}{RTS}{ray tracing simulator}
\newacronym{ru}{RU}{Radio Unit}
\newacronym{rv}{RV}{random variable}
\newacronym{rw}{RW}{RadioWeaves}
\newacronym{rx}{RX}{receiver}
\newacronym{rzf}{RZF}{regularized zero forcing}
\newacronym{s-parameter}{S-parameter}{scattering parameter}
\newacronym{sa}{SA}{synchronization anchor}
\newacronym{sa5}{SA}{Stand Alone}
\newacronym{sar}{SAR}{specific absorption rate}
\newacronym{sbl}{SBL}{sparse Bayesian learning}
\newacronym{sc}{SC}{single carrier}
\newacronym{scfde}{SC-FDE}{single-carrier modulation with frequency-domain-equalization}
\newacronym{scfdma}{SCFDMA}{single-carrier frequency division multiple access}
\newacronym{scs}{SCS}{sub-carrier spacing}
\newacronym{sdg}{SDG}{Sustainable Development Goal}
\newacronym{sdm}{SDM}{sigma-delta modulator}
\newacronym{sdma}{SDMA}{spatial-division multiple access}
\newacronym{sdn}{SDN}{software-defined network}
\newacronym{sdof}{SDoF}{sigma-delta over fiber}
\newacronym{sdr}{SDR}{software-defined radio}
\newacronym{se}{SE}{spectral efficiency}
\newacronym{sei}{SEI}{specific emitter identification}
\newacronym{ser}{SER}{symbol-error rate}
\newacronym{sf}{SF}{spreading factor}
\newacronym{sfn}{SFN}{single frequency network}
\newacronym{sfp}{SFP}{small form-factor pluggable}
\newacronym{sha}{SHA}{Secure Hash Algorithm}
\newacronym{SigMF}{SigMF}{Signal Metadata Format}
\newacronym{simo}{SIMO}{single-input multiple-output}
\newacronym{sinr}{SINR}{signal-to-interference-plus-noise ratio}
\newacronym{siso}{SISO}{single-input single-output}
\newacronym{slam}{SLAM}{simultaneous localization and mapping}
\newacronym{slc}{SLC}{spatial leakage suppression}
\newacronym{slerp}{SLERP}{spherical linear interpolation}
\newacronym{sma}{SMA}{SubMiniature version A}
\newacronym{smc}{SMC}{specular multipath component}
\newacronym{smps}{SMPS}{switched mode power supply}
\newacronym{sndr}{SNDR}{signal-to-noise-and-distortion ratio}
\newacronym{snidr}{SNIDR}{signal-to-noise-and-interference-and-distortion ratio}
\newacronym{snir}{SNIR}{signal-to-interference-plus-noise ratio}
\newacronym{snr}{SNR}{signal-to-noise ratio}
\newacronym{soc}{SoC}{state of charge}
\newacronym{SoC}{SOC}{System on Chip}
\newacronym{sp}{SP}{service point}
\newacronym{spdt}{SPDT}{single pole double throw}
\newacronym{spi}{SPI}{Serial Peripheral Interface}
\newacronym{spst}{SPST}{single pole single throw}
\newacronym{sram}{SRAM}{static random-access memory}
\newacronym{srd}{SRD}{short-range device}
\newacronym{srls}{SRLS}{standard recursive least squares}
\newacronym{srs}{SRS}{Sounding Reference Signal}
\newacronym{ssb}{SSB}{synchronisation signal block}
\newacronym{ssd}{SSD}{solid state drive}
\newacronym{ssq}{SSQ}{simulator sickness questionnaire}
\newacronym{steam}{STEAM}{science, technology, engineering, the arts, and mathematics}
\newacronym{svd}{SVD}{singular value decomposition}
\newacronym{sw}{SW}{spherical wavefront}
\newacronym{swipt}{SWIPT}{simultaneous wireless information and power transfer}
\newacronym{synce}{SyncE}{Synchronous Ethernet}
\newacronym{ta}{TA}{timing advance}
\newacronym{tau}{TAU}{tracking area update}
\newacronym{tcer}{TCER}{transported to consumed energy ratio}
\newacronym{tcp}{TCP}{Transmission Control Protocol}
\newacronym{tcxo}{TCXO}{temperature compensated crystal oscillator}
\newacronym{tdce}{TD-CE}{time-domain compression and expansion}
\newacronym{tdd}{TDD}{time division duplexing}
\newacronym{tdma}{TDMA}{time division-multiple access}
\newacronym{tdoa}{TDOA}{time-difference-of-arrival}
\newacronym{thz}{THz}{Terahertz}
\newacronym{to}{TO}{timing offset}
\newacronym{toa}{TOA}{time-of-arrival}
\newacronym{tof}{ToF}{time-of-flight}
\newacronym{tosm}{TOSM}{through-open-short-match}
\newacronym{tpms}{TPMS}{Tire-Pressure Monitoring System}
\newacronym{trl}{TRL}{technology readyness level}
\newacronym{trp}{TRP}{Transmission Reception Point}
\newacronym{tsn}{TSN}{time-sensitive networking}
\newacronym{ttf}{TTF}{tag talks first}
\newacronym{ttff}{TTFF}{Time To First Fix}
\newacronym{tti}{TTI}{transmission time interval}
\newacronym{ttm}{TTM}{time to market}
\newacronym{ttn}{TTN}{The Things Network}
\newacronym{tx}{TX}{transmitter}
\newacronym{uart}{UART}{Universal Asynchronous Receiver/Transmitter}
\newacronym{uav}{UAV}{unmanned aerial vehicle}
\newacronym{uc}{UC}{use case}
\newacronym{ucie}{UCIe}{Universal Chiplet Interconnect Express}
\newacronym{udp}{UDP}{User Datagram Protocol}
\newacronym{ue}{UE}{user equipment}
\newacronym{ugv}{UGV}{unmanned ground vehicle}
\newacronym{uhd}{UHD}{USRP hardware driver}
\newacronym{uhf}{UHF}{ultra-high frequency}
\newacronym{ul}{UL}{uplink}
\newacronym{ula}{ULA}{uniform linear array}
\newacronym{uMIMO}{$\mu$-MIMO}{ultra-massive Multiple-Input Multiple-Output}
\newacronym{ummtc}{umMTC}{ultra massive machine type communication}
\newacronym{un}{UN}{United Nations}
\newacronym{unow}{uNOW}{unified non-orthogonal waveform}
\newacronym{upa}{UPA}{uniform planar array}
\newacronym{ura}{URA}{uniform rectangular array}
\newacronym{urllc}{URLLC}{ultra-reliable low-latency communications}
\newacronym{usrp}{USRP}{universal software radio peripheral}
\newacronym{uv}{UV}{unmanned vehicle}
\newacronym{uw}{UW}{unique word}
\newacronym{uwb}{UWB}{ultrawideband}
\newacronym{uwofdm}{UW-DFT-s-OFDM}{unique word DFT-s-OFDM}
\newacronym{v2v}{V2V}{vehicle-to-vehicle}
\newacronym{vco}{VCO}{voltage-controlled oscillator}
\newacronym{vep}{VEP}{virtual edge platform}
\newacronym{vlc}{VLC}{visible light communication}
\newacronym{vlp}{VLP}{visible light positioning}
\newacronym{vna}{VNA}{vector network analyzer}
\newacronym{voc}{VOC}{Voltatile Organic Compound}
\newacronym{votable}{VOTable}{Virtual Observatory Table}
\newacronym{vr}{VR}{virtual reality}
\newacronym{vuca}{VUCA}{volatile, uncertain, complex and ambiguous}
\newacronym{wb}{WB}{wideband}
\newacronym{wban}{WBAN}{wireless body area network}
\newacronym{wd}{WD}{Wireless distance}
\newacronym{wimax}{WiMAX}{Worldwide Interoperability for Microwave Access}
\newacronym{wlan}{WLAN}{wireless LAN}
\newacronym[plural=WPs,firstplural=work packages (WPs)]{wp}{WP}{work package}
\newacronym{wpt}{WPT}{wireless power transfer}
\newacronym{wr}{WR}{White Rabbit}
\newacronym{wrsn}{WRSN}{wireless rechargeable sensor network}
\newacronym{wsn}{WSN}{wireless sensor network}
\newacronym{xets}{XETS}{cross exponentially tapered slot}
\newacronym{xlmimo}{XL-MIMO}{extremely large-scale MIMO}
\newacronym{xr}{XR}{extended reality}
\newacronym{z3ro}{Z3RO}{zero third-order distortion}
\newacronym{zf}{ZF}{zero-forcing}
\newacronym{zmcscg}{ZMCSCG}{zero mean circularly symmetric complex Gaussian}
\newacronym{zmq}{ZMQ}{ZeroMQ}
\newacronym{ztofdm}{ZT-DFT-s-OFDM}{zero-tail DFT-s-OFDM}
\newacronym{llr}{LLR}{log-likelihood ratio}
\newacronym{rps}{RPS}{random-phase sweeping}

\input{auto_names}                  %

\usepackage{orcidlink}

\usepackage[backend=biber,style=ieee]{biblatex}
\AtBeginBibliography{\footnotesize}

\def\BibTeX{{\rm B\kern-.05em{\sc i\kern-.025em b}\kern-.08em
    T\kern-.1667em\lower.7ex\hbox{E}\kern-.125emX}}

\definecolor{color1}{HTML}{1b9e77}
\definecolor{color2}{HTML}{d95f02}
\definecolor{color3}{HTML}{7570b3}
\definecolor{color4}{HTML}{e7298a}
\definecolor{color5}{HTML}{66a61e}
\definecolor{color6}{HTML}{e6ab02}
\definecolor{color7}{HTML}{a6761d}
\definecolor{color8}{HTML}{666666}

\colorlet{AMBgreen}{color1}
\colorlet{AMBdarkgreen}{color1}
\colorlet{AMBlightgreen}{color1}
\definecolor{darkgray176}{RGB}{176,176,176}

\usepackage{tikz}
\usepackage{pgfplots}   %
\pgfplotsset{compat=newest}
\usetikzlibrary{plotmarks}
\usetikzlibrary{arrows.meta}
\usetikzlibrary{arrows}
\usetikzlibrary{calc,fadings,decorations.pathreplacing}
\usetikzlibrary{fit,backgrounds}  %

\usetikzlibrary{shapes,arrows,fit,positioning}
\usetikzlibrary{shapes}
\usetikzlibrary{spy}
\usetikzlibrary{circuits}
\usetikzlibrary{arrows}

\pgfplotsset{
    every axis/.append style={
        title style={draw=none},
        label style={font=\small},
        legend style={
            fill opacity=0.8,
            nodes={scale=0.8, transform shape}, {draw=none}
        },
        tick align=outside,
        tick pos=left,
        x grid style={darkgray176},
        xtick style={color=black},
        y grid style={darkgray176},
        ytick style={color=black},
        grid=both,
    },
    every axis plot/.append style={
        line width=2.0pt,
    },
}

\tikzset{%
  >=latex,
  inner sep=0pt,%
  outer sep=2pt,%
  mark coordinate/.style={inner sep=0pt,outer sep=0pt,minimum size=3pt,
  fill=black,circle}%
}

\pgfplotsset{colormap={inferno}{
    rgb(0)=(0.001462, 0.000466, 0.013866),
    rgb(15)=(0.037668, 0.025921, 0.132232),
    rgb(30)=(0.116656, 0.047574, 0.272321),
    rgb(45)=(0.217949, 0.036615, 0.383522),
    rgb(60)=(0.316282, 0.053490, 0.425116),
    rgb(75)=(0.410113, 0.087896, 0.433098),
    rgb(90)=(0.503493, 0.121575, 0.423356),
    rgb(105)=(0.596940, 0.154848, 0.398125),
    rgb(120)=(0.688653, 0.192239, 0.357603),
    rgb(135)=(0.775059, 0.239667, 0.303526),
    rgb(150)=(0.851384, 0.302260, 0.239636),
    rgb(165)=(0.912966, 0.381636, 0.169755),
    rgb(180)=(0.956852, 0.475356, 0.094695),
    rgb(195)=(0.981895, 0.579392, 0.026250),
    rgb(210)=(0.987464, 0.690366, 0.079990),
    rgb(225)=(0.973088, 0.805409, 0.216877),
    rgb(240)=(0.947594, 0.917399, 0.410665),
    rgb(255)=(0.988362, 0.998364, 0.644924),
    }}

\colorlet{scenarioColor1}{color8}
\colorlet{scenarioColor2}{color8}

\colorlet{strategyColorGEO}{color1}
\colorlet{strategyColorCSI}{color2}
\colorlet{strategyColorRPS}{color3}

\newcommand{\scenarioTag}[1]{\roundLabel[scenarioColor#1]{Scenario~#1}}

\newcommand{\CSI}{\roundLabel[strategyColorCSI]{\glsentryshort{csi}}\xspace}
\newcommand{\GEO}{\roundLabel[strategyColorGEO]{GEO}\xspace}
\newcommand{\RPS}{\roundLabel[strategyColorRPS]{\glsentryshort{rps}}\xspace}

\newcommand{\csi}{\glsentryshort{csi}\xspace}
\newcommand{\geo}{geometry\xspace}
\newcommand{\rps}{\glsentryshort{rps}\xspace}

\usepackage[font=small]{caption}
\usepackage{subcaption}

\usepackage{fontawesome}
\usetikzlibrary {patterns,patterns.meta}

\defineauthors[false]{gilles, jarne}

\begin{document}

\title{Experimental Evaluation of Geometry and Reciprocity-Based Beamforming with Large Arrays for RF Wireless Power Transfer
\thanks{This work was supported by the AMBIENT-6G project, which received funding from the Smart Networks and Services Joint Undertaking (SNS JU) under the European Union's Horizon Europe research and innovation programme under Grant Agreement No. 101192113, and by the Research Foundation - Flanders (FWO) through a Junior Postdoctoral Fellowship, project “Cocoon: Towards Fluid Energy-Efficient Open Access Networks through Citizen’s Co-Creation” (grant/application no. 12A2V25N).}
\thanks{For the purpose of open access, the author has applied a CC BY public copyright license to any Author Accepted Manuscript version arising from this submission.}
\thanks{This paper is accepted at IEEE Wireless Power Technologies Conference and Expo 2026 (WPTCE 2026) and can be found in~\cite{Call2607:Experimental}.}
}

\author{
    \IEEEauthorblockN{%
    Gilles Callebaut, Jarne Van Mulders}%
    \IEEEauthorblockA{\textit{Department of Electrical Engineering, KU Leuven}, Belgium}
}

\maketitle

\begin{abstract}
This paper experimentally investigates geometry-based multi-antenna \gls{rf} \gls{wpt} using a large-scale \SI{8}{\meter}~$\times$~\SI{4}{\meter} distributed indoor transmit array. Geometry-based beamforming exploits known transmitter–receiver positions to perform phase-only precoding, avoiding explicit channel estimation or feedback. Using a ceiling-mounted array of \num{41} phase-synchronized transmit antennas operating at \SI{920}{\mega\hertz}, we compare geometry-based beamforming against \gls{csi}-based beamforming. Spatial power delivery is evaluated through two-dimensional scans over a \SI{1.25}{\meter}~$\times$~\SI{1.25}{\meter} area, with harvested \gls{dc} power measured using an RF-to-DC energy profiler. 
Under \gls{los} conditions, geometry-based beamforming achieves a power gain of \SI{18.75}{\decibel}, within \SI{0.82}{\decibel} of \gls{csi}-based beamforming. In obstructed \gls{los} scenarios with reflections, its gain reduces to \SI{16.7}{\decibel}, while \gls{csi}-based beamforming maintains \SI{20.53}{\decibel}, resulting in a \SI{3.83}{\decibel} performance gap. These results quantify the trade-off between reduced system overhead and robustness to multipath propagation in geometry-driven \gls{wpt}, and constitute a first step toward geometry-based \gls{wpt} enabled by digital twins.
\end{abstract}


\section{Introduction}

\glsresetall

Radiative \gls{rf} \gls{wpt} has emerged as a key enabling technology for batteryless and maintenance-free systems in the context of the Internet of Things, embedded sensing, and cyber--physical infrastructures. By exploiting electromagnetic radiation in the radiative near field and far field, \gls{rf} \gls{wpt} enables energy delivery over meter-scale distances without physical contact, overcoming the severe range limitations of inductive or resonant near-field techniques~\cite{ClerckxFoundations2022}.

A fundamental limitation of \gls{rf} \gls{wpt} is the inherently low end-to-end power transfer efficiency, dominated by free-space path loss. Multi-antenna transmission and beamforming are therefore central tools to spatially concentrate radiated energy at a target receiver. Prior experimental work has demonstrated substantial power gains using coherent transmission from collocated or distributed antenna arrays~\cite{EnergyBall2018,ShenDAS2021}. Existing experimental platforms can broadly be categorized based on the beamforming strategy employed and the degree of synchronization between distributed transmitters.

A first class of systems adopts \emph{geometry-based deterministic beamforming}, where transmit weights are computed from explicit propagation models using known antenna and receiver positions. This approach is commonly based on conjugate-matched beamforming, assuming either far-field plane-wave propagation or near-field spherical-wave models. Such techniques have been analyzed theoretically and recently validated experimentally in centralized and distributed antenna configurations~\cite{ClerckxFoundations2022,ShenDAS2021,9270598,10000769,deutschmann2025physically}. 

A second class of systems relies on \emph{distributed adaptive beamforming} using receiver-side power feedback. Representative examples include Energy-Ball--type systems, where multiple spatially separated transmitters iteratively adjust their phases using low-rate or one-bit feedback to maximize the harvested power~\cite{EnergyBall2018}. These approaches do not require explicit channel estimation or geometric knowledge and can operate without wired synchronization. However, convergence is stochastic, slow for large arrays, and highly sensitive to noise and mobility, limiting scalability and reproducibility.

Related work has also explored \emph{blind or feedback-driven beamforming} techniques, where phase perturbations are applied and accepted or rejected based on monotonic power measurements at the receiver~\cite{YedavalliBlindBeamforming2017}. While conceptually simple, these methods typically exhibit poor convergence properties and are unsuitable for large-scale arrays or high-resolution spatial power control.

More recent studies have investigated \emph{concurrent and asynchronous \gls{wpt}}, where multiple transmitters radiate energy without explicit phase alignment~\cite{UnderstandingConcurrentWPT2024}. Although such systems simplify deployment and eliminate synchronization requirements, they inherently forgo coherent beamforming gains and therefore suffer from significantly reduced energy efficiency.

Across these works, synchronization emerges as a critical differentiator. Geometry-based and near-field beamforming experiments almost exclusively rely on centralized clock distribution and phase-coherent radio front-ends to guarantee stable phase relationships across antennas~\cite{ShenDAS2021,ClerckxFoundations2022}. In contrast, feedback-based systems implicitly achieve phase alignment through iterative adaptation, while asynchronous systems accept non-coherent power combining as a design trade-off~\cite{MittalClosedLoop2024}.

\subsection{Toward Digital Twin–Enabled \gls{wpt}.}
Recent advances in wireless systems have highlighted the potential of \emph{digital twins}---virtual replicas of physical environments enriched with geometric, material, and electromagnetic information---to support environment-aware communication and sensing. In particular, geometry-driven channel modeling using ray tracing and site-specific simulations has been shown to accurately predict propagation characteristics in complex indoor environments and across frequency bands, forming the basis for digital-twin-assisted beamforming and network optimization~\cite{11115919}.

While digital twin concepts are gaining traction in wireless communications, their application to \gls{rf} \gls{wpt} remains largely unexplored experimentally. Existing \gls{wpt} platforms either rely on idealized free-space models, purely feedback-driven adaptation, or implicit exploitation of multipath through channel reciprocity. Only a limited number of recent works have begun to integrate environment sensing and geometry-aware control into \gls{wpt} systems, for example by using position information or environmental mapping to steer energy beams and enforce safety constraints~\cite{11037283}.

As a result, a systematic experimental assessment of geometry-based beamforming for \gls{wpt} in realistic indoor environments, and a quantitative comparison against \gls{csi}-based approaches, is still missing. Bridging this gap is a necessary first step toward digital twin–enabled \gls{wpt}, where beamforming decisions are derived from a virtual representation of the environment rather than from continuous receiver feedback or full channel sounding.

This work addresses this gap by experimentally evaluating geometry-based multi-antenna \gls{rf} \gls{wpt} using a large-scale distributed indoor array and by directly comparing its performance against \gls{csi}-based beamforming under both \gls{los} and \gls{olos} conditions. The presented results quantify the performance loss due to unmodeled multipath propagation and establish an experimental baseline for future geometry-driven and digital twin–assisted \gls{wpt} systems~\cite{9814679,10283480}.

\begin{figure*}[ht!]
    \centering
    \begin{subfigure}[t]{0.43\linewidth}
        \centering
        \includegraphics[height=6.2cm,trim={15cm 0 15cm 0},clip]{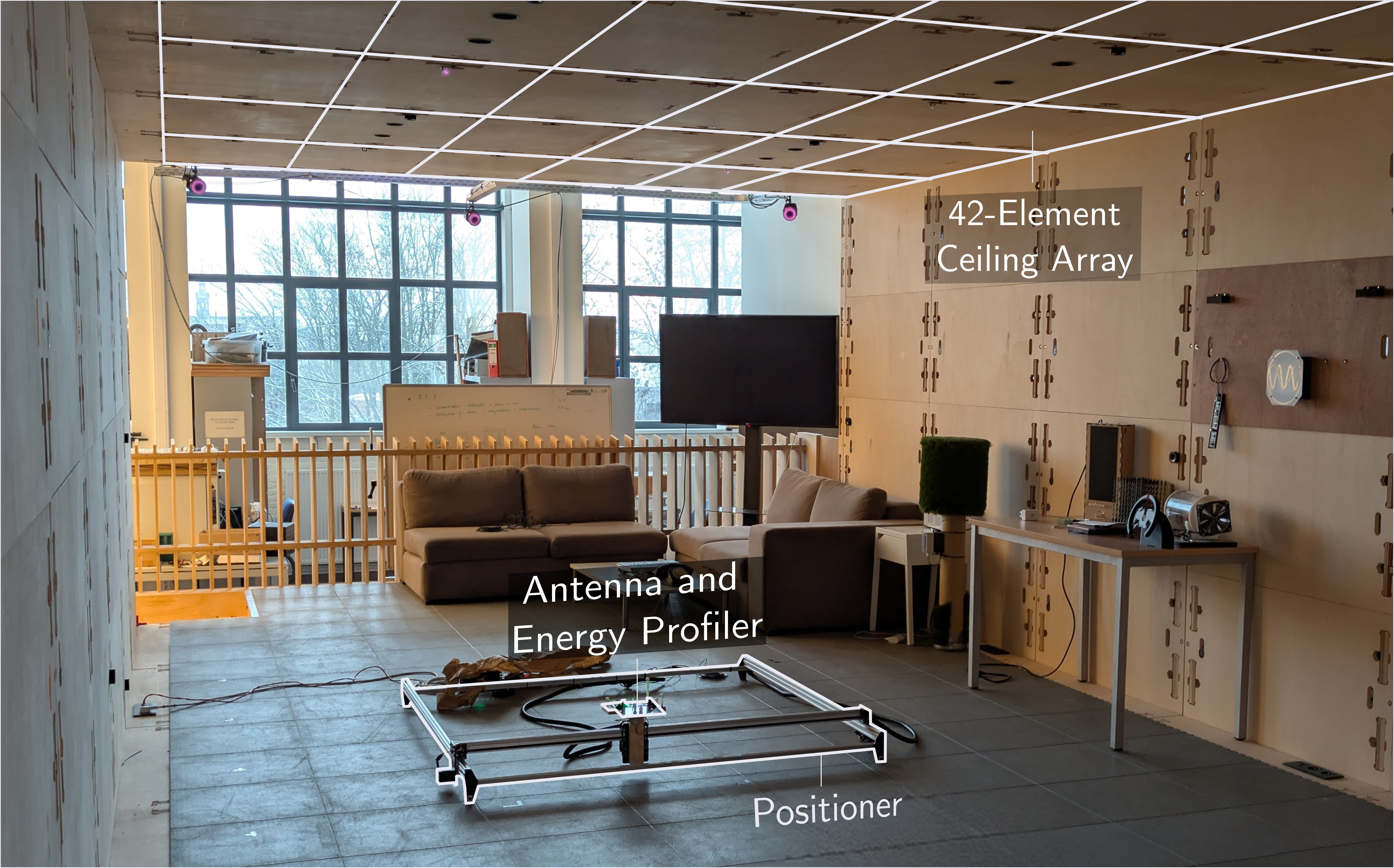}
        \caption{\scenarioTag{1} Energy profiler, XY plotter, and ceiling-integrated transmit array.}
        \label{fig:setup:overview}
    \end{subfigure}\hspace*{0.2cm}
    \begin{subfigure}[t]{0.36\linewidth}
        \centering
        \includegraphics[height=6.2cm]{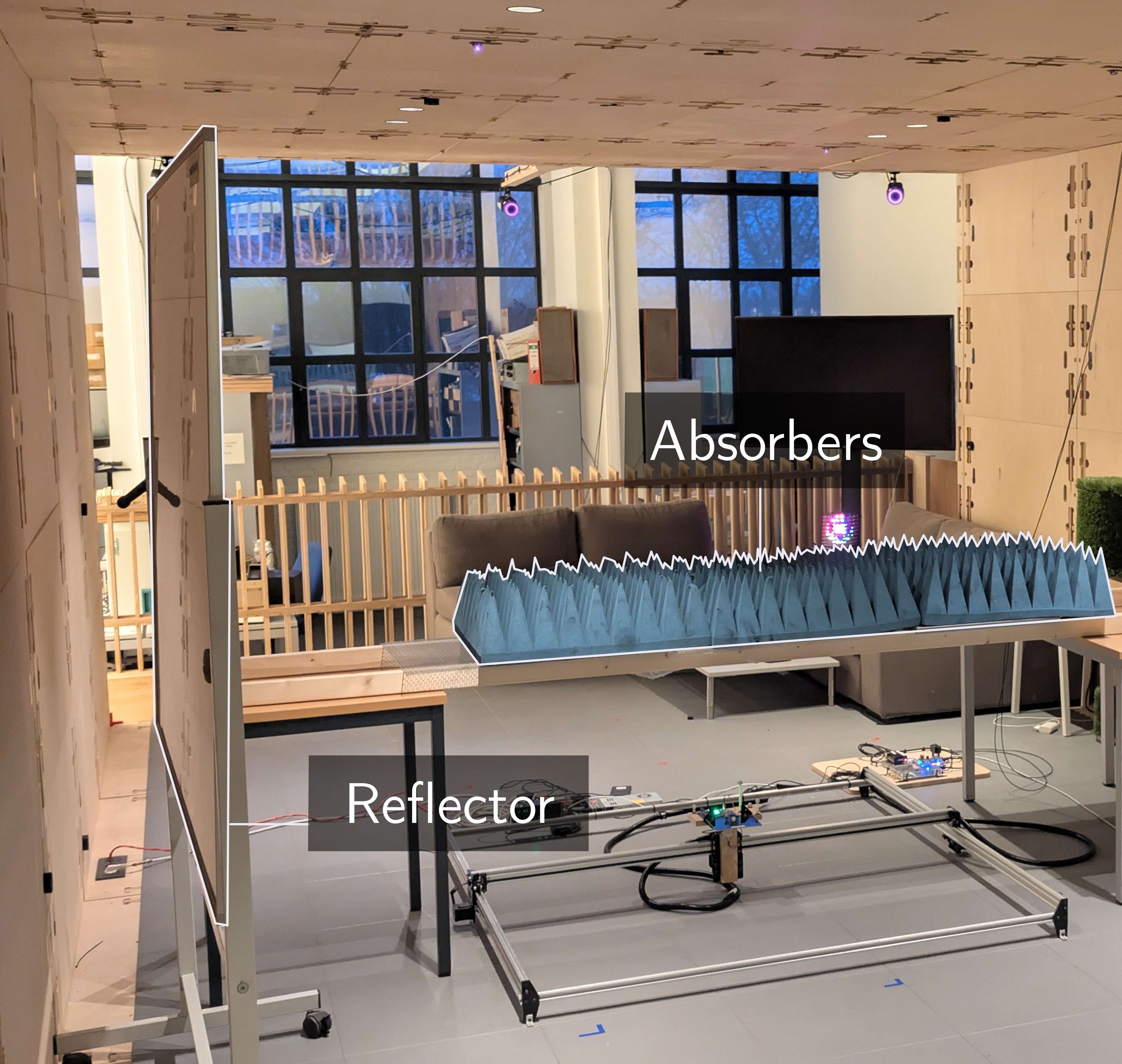}
        \caption{\scenarioTag{2} \Gls{olos} scenario with reflector and absorbers.}
        \label{fig:setup:olos}
    \end{subfigure}\hspace*{0.2cm}
    \begin{subfigure}[t]{0.17\linewidth}
        \centering
        \includegraphics[height=6.2cm]{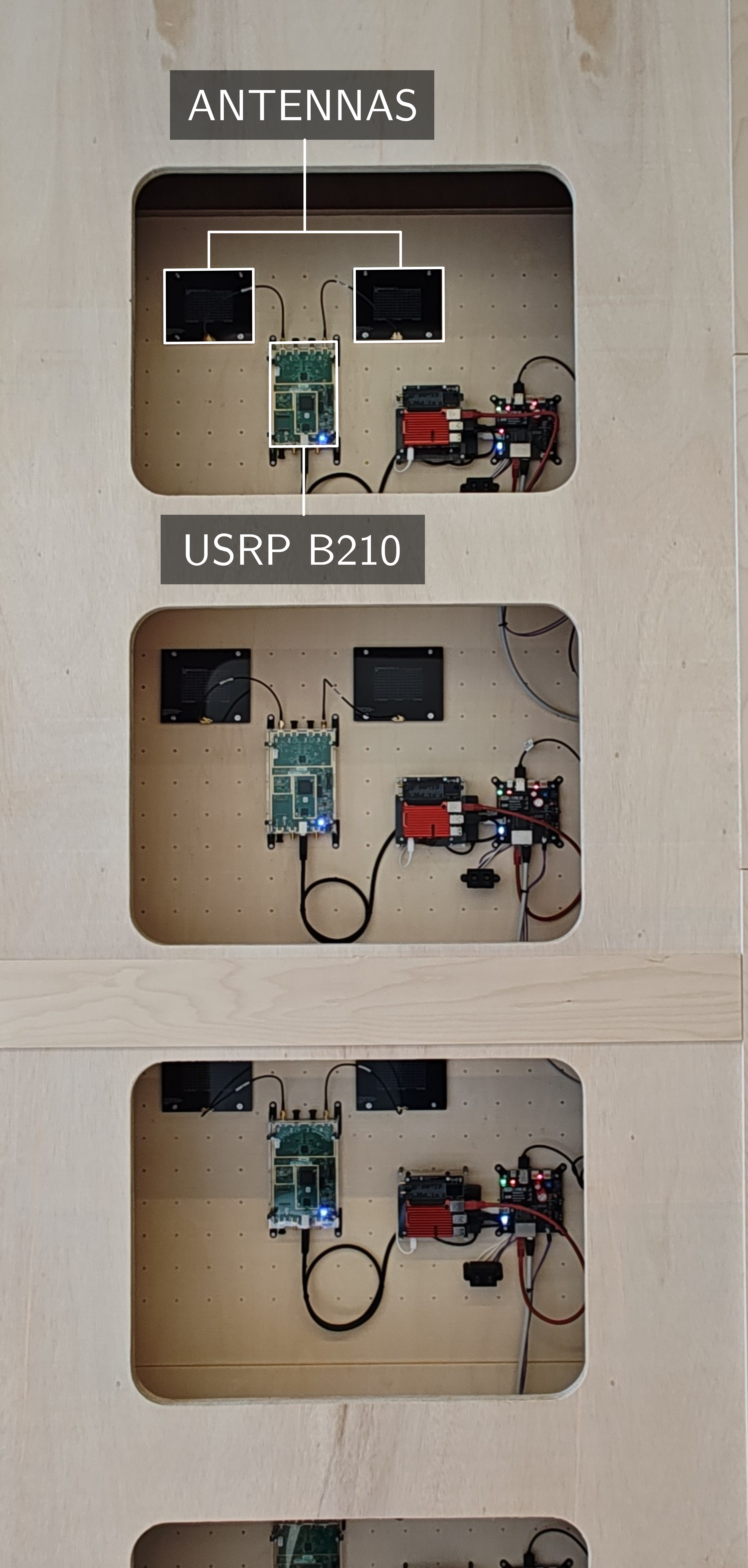}
        \caption{Close-up of TX array elements.}
        \label{fig:setup:elements}
    \end{subfigure}
    \caption{Experimental setup.}
    \label{fig:setup}
\end{figure*}

\subsection{Application Scenarios}

Geometry-based multi-antenna \gls{rf} \gls{wpt} is well suited to environments where transmitter and receiver positions are fixed or accurately known. Typical applications include indoor batteryless sensing with ceiling-mounted infrastructure enabling selective node energization~\cite{ShenDAS2021}, instrumented laboratory and robotic environments with precise real-time position tracking~\cite{EnergyBall2018}, and smart buildings where largely static geometry and dominant reflections allow predictable and reproducible power delivery~\cite{ClerckxFoundations2022}.

\subsection{Contributions}

The main contributions of this work are summarized as follows:
\begin{itemize}
    \item We present a large-scale experimental evaluation of geometry-based multi-antenna \gls{rf} \gls{wpt} using a ceiling-mounted distributed array of \num{41} phase-synchronized transmit antennas operating at \SI{920}{\mega\hertz}.
    \item We quantitatively compare geometry-based beamforming against \gls{csi}-based beamforming under both \gls{los} and \gls{olos} propagation conditions, using spatially resolved two-dimensional power scans.
    \item We experimentally show that geometry-based beamforming achieves within \SI{0.82}{\decibel} of \gls{csi}-based performance under \gls{los} conditions, while exhibiting a \SI{3.83}{\decibel} performance loss in reflective \gls{olos} environments.
    \item We provide experimental evidence of the impact of multipath propagation on geometry-driven phase-only precoding and identify the performance gap that arises when deterministic reflections are not modeled.
    \item We establish an experimental baseline for future geometry-based \gls{wpt} approaches incorporating digital twins or hybrid geometric–measurement-based beamforming.
    \item All source code and data is publicly available in~\cite{github}~\faicon{github}.
\end{itemize}

\section{Experimental Setup and Measurement Methodology}

The experiments were conducted using the Techtile distributed \gls{mimo} testbed~\cite{Call2206:Techtile}. The setup is shown in~\cref{fig:setup}. 
It consists of \num{42} ceiling-mounted active antenna elements\footnote{All experiments were conducted with \num{41} antennas due to a faulty USRP.}  arranged in a fixed grid. Each element integrates a custom-designed patch antenna with characterized radiation properties and a USRP B210. Phase coherence across the array is ensured through the calibration procedure as detailed below. Beamforming weights are computed and applied using a \gls{uhd}-based Python control framework. 

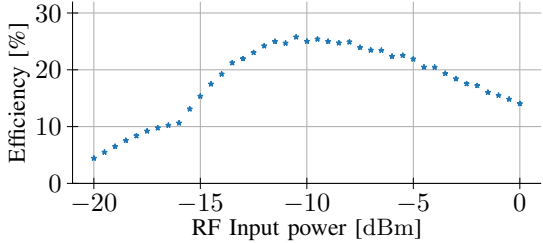
\begin{figure}
    \centering
    \begin{tikzpicture}

\definecolor{crimson2143940}{RGB}{214,39,40}
\definecolor{darkgray176}{RGB}{176,176,176}
\definecolor{darkorange25512714}{RGB}{255,127,14}
\definecolor{darkturquoise23190207}{RGB}{23,190,207}
\definecolor{forestgreen4416044}{RGB}{44,160,44}
\definecolor{goldenrod18818934}{RGB}{188,189,34}
\definecolor{gray127}{RGB}{127,127,127}
\definecolor{lightgray204}{RGB}{204,204,204}
\definecolor{mediumpurple148103189}{RGB}{148,103,189}
\definecolor{orchid227119194}{RGB}{227,119,194}
\definecolor{sienna1408675}{RGB}{140,86,75}
\definecolor{steelblue31119180}{RGB}{31,119,180}

\begin{axis}[
width=0.9\linewidth,
height=4cm,
xlabel={RF Input power [\si{\dBm}]},
ylabel={Efficiency [\%]},
xmin=-21, xmax=1,
ymin=0, ymax=32,
axis lines=left,
axis line style={-},
grid=both,
grid style={darkgray176},
legend cell align={left},
legend style={at={(0.03,0.97)}, anchor=north west, draw=lightgray204, fill opacity=0.8, text opacity=1},
]
\addplot [only marks, mark=star, mark size=0.5pt, steelblue31119180]
table {%
-20 4.43
-19.5 5.49010577890387
-19 6.50554824239186
-18.5 7.56794043506644
-18 8.37280396125217
-17.5 9.21115090661792
-17 9.76312731105926
-16.5 10.2379879321001
-16 10.6254803820728
-15.5 13.0961621966113
-15 15.3496957624573
-14.5 17.4908844714272
-14 19.2435619517949
-13.5 21.2185989513507
-13 21.9478854646577
-12.5 23.0322749188241
-12 24.1997341556887
-11.5 24.9877891643765
-11 24.6887862506954
-10.5 25.7693978399532
-10 25.017
-9.5 25.4006517368117
-9 24.9967152185384
-8.5 24.7179270617722
-8 24.895683941155
-7.5 23.9326844587761
-7 23.4580684699245
-6.5 23.4258743067651
-6 22.3656608416955
-5.5 22.5292309627751
-5 21.8911833302816
-4.5 20.4685060383081
-4 20.4470067411311
-3.5 19.352624882354
-3 18.40390054081
-2.5 17.544860315326
-2 17.2119400701277
-1.5 16.0243909212184
-1 15.5041700164099
-0.5 14.8199563499566
0 14.0402
};
\end{axis}

\end{tikzpicture}
    \caption{End-to-end efficiency of the \gls{ep} at \SI{925}{\mega\hertz}, measured for a configured output DC target voltage of \SI{1.62}{\volt} at the energy harvester. This voltage level is sufficient to power most low-power microcontrollers. The \gls{rf} input power represents the \gls{rf} signal power at \SI{925}{\mega\hertz} that is directly applied to the input of the \gls{ep}. The obtained efficiencies are in line with literature.}\label{fig:end-efficiency}
\end{figure}

On the receiving side, harvested energy is measured using a custom-designed compact \gls{rf}-to-\gls{dc} rectification unit~\cite{10765709}, referred to as the \emph{Energy Profiler}. The \gls{ep} integrates a tuning network for impedance matching to \SI{50}{\ohm} antennas and an RF energy harvester delivering a \gls{dc} output voltage, constrained to a maximum of \SI{2}{\volt}. Precise power monitoring is enabled via a programmable potentiometer and a \num{16}-bit \gls{adc}. Operating with a \SI{1}{\kilo\hertz} sampling and control loop, the \gls{ep} supports fast adaptation in dynamic scenarios. The conversion efficiency is depicted in \cref{fig:end-efficiency}. 

To support channel state information acquisition, the \gls{ep} is equipped with a circulator that enables dual operation. During the pilot transmission phase, the antenna is connected to a USRP B210, allowing the target node to transmit a pilot signal from the exact energy harvesting location, as the same antenna is used. In the subsequent energy transfer phase, incoming \gls{rf} signals are routed through the circulator to the rectification circuitry.

To map the spatial distribution of delivered power, the receiver is mounted on a motorized XY-plotter that scans a two-dimensional grid within the measurement area of \SI{1.25}{\meter}~$\times$~\SI{1.25}{\meter}. 

Ground-truth positioning is obtained using a Qualisys motion capture system. Retro-reflective markers attached to the receiver enable three-dimensional tracking with millimeter-level accuracy. The Qualisys coordinate frame is aligned with the Techtile reference frame to ensure consistency between measured positions and geometric beamforming models.

Phase coherence across all \num{42} transmit channels is achieved using a multi-step calibration pipeline~\faicon{github}\footnote{\url{https://github.com/techtile-by-dramco/NI-B210-Sync}}:
\begin{enumerate}
\item Injection of a reference \gls{rf} tone to establish an absolute phase anchor.
\item Loopback calibration to correct TX--RX asymmetries in the \gls{lo} paths.
\item Compensation of static phase offsets introduced by reference cables.
\end{enumerate}

\subsection{Scenarios}
\label{sec:scenarios}

\subsubsection{\texorpdfstring{\scenarioTag{1}}{Scenario 1} (\gls{los})}
\label{sec:scenario1}
The baseline measurement scenario features an unobstructed propagation path between the ceiling-mounted transmit array and the receiver over the entire scan area (\cref{fig:setup:overview}).
It is intended to approximate free-space propagation, such that deterministic, geometry-based phase alignment can be expected to perform close to \csi{}-based beamforming.

\subsubsection{\texorpdfstring{\scenarioTag{2}}{Scenario 2} (\gls{olos} with reflector and absorbers)}
\label{sec:scenario2}
To study robustness against multipath and model mismatch, an \gls{olos} scenario is created by obstructing the direct path through absorbers and introducing a strong specular reflection using a reflector (\cref{fig:setup:olos}).
This controlled reflective setup is used throughout the experimental results to quantify the performance degradation of geometry-only precoding in the presence of reflections.

\subsection{Beamforming Strategies}

We consider three beamforming strategies. \rps{} is used as a non-coherent baseline, while \csi{} and \geo{}-based beamforming aim to coherently combine power at the intended receiver location. Throughout this work, only the phase of the transmit weight $w_m$ is applied, such that all antennas radiate equal average RF power.

\subsubsection{\texorpdfstring{\CSI}{CSI}-Based Beamforming}
In \csi{}-based beamforming, the transmit weights are derived from an estimate of the complex baseband channel between each transmit antenna and the receiver. Let $h_m \in \mathbb{C}$ denote the complex baseband channel coefficient between transmit antenna $m$ and the receiver, estimated via channel reciprocity. The transmit weight is obtained by conjugate matching,
\begin{equation}
    w_m = \hat{h}_m^{*}.
\end{equation}
Under perfect channel knowledge, $\hat{h}_m = h_m$, all transmitted signals add coherently at the receiver, yielding a received power scaling proportional to $M^2$ relative to single-antenna transmission.

\subsubsection{\texorpdfstring{\GEO}{GEO}-Based Beamforming}
Geometry-based beamforming avoids explicit channel estimation by exploiting known antenna and receiver positions. Let $\mathbf{p}_m \in \mathbb{R}^3$ denote the position of transmit antenna $m$ and $\mathbf{r}_{\mathrm{f}} \in \mathbb{R}^3$ the target focal point. Assuming free-space line-of-sight propagation and unit-amplitude channels, the channel phase is fully determined by geometry,
\begin{equation}
    h_m^{\mathrm{geo}} = e^{j k d_m},
\end{equation}
where $d_m = \lVert \mathbf{p}_m - \mathbf{r}_{\mathrm{f}} \rVert$ and $k = 2\pi / \lambda$.
This strategy enables coherent combining without CSI feedback, but its accuracy depends on positioning precision and the validity of the line-of-sight assumption.

\subsubsection{\texorpdfstring{\RPS}{RPS} (Random-Phase Sweeping)}

As a non-coherent baseline, each transmit antenna applies an independent random phase $\phi_m \sim \mathcal{U}(0,2\pi)$, updated every \SI{100}{\milli\second}, such that $x_m = e^{j\phi_m}$. Due to the independence of the phases, all cross-terms vanish in expectation, yielding an average received power
\begin{equation}
    P_{\mathrm{rx,non\text{-}coh}}
    = \sum_{m=1}^{M} \mathbb{E}[|h_m|^2] \, P_1 .
\end{equation}
Assuming identical average channel gains, this simplifies to
\begin{equation}
    P_{\mathrm{rx,non\text{-}coh}} = M \, \mathbb{E}[|h|^2] \, P_1 ,
\end{equation}
corresponding to a linear power scaling of $M$ relative to single-antenna transmission.

\section{Experimental Results and Analysis}

This section reports the experimental evaluation of the proposed beamforming strategies and analyzes their spatial power delivery characteristics. The experimental parameters are summarized in \cref{tab:exp_params}. All measurements report harvested \gls{dc} power at the receiver, accounting for the measured RF-to-DC conversion efficiency.

\begin{table}[t]
\centering
\caption{Experimental parameters}
\label{tab:exp_params}
\begin{tabular}{ll}
\toprule
Parameter & Value \\
\midrule
TX transmit power per antenna & \SI{11}{\dBm} \\
Number of TX antennas ($M$) & \num{41} \\
Total TX transmit power & \SI{27}{\dBm} \\
TX antenna type & Patch antenna \\
RX antenna type & Patch antenna \\
Carrier frequency ($f_\mathrm{c}$) & \SI{920}{\mega\hertz} \\
Grid scan area & \SI{1.25}{\meter} $\times$ \SI{1.25}{\meter} \\
RF-to-DC efficiency & \cref{fig:end-efficiency} \\
\bottomrule
\end{tabular}
\end{table}

\subsection{Wireless Power Heatmaps}

\begin{figure*}[t]
    \centering
    \newlength{\heatmapheight}
    \setlength{\heatmapheight}{3.8cm}
    \begin{subfigure}[t]{0.3\linewidth}
        \centering
        \resizebox{!}{\heatmapheight}{%
            \begin{tikzpicture}
  \begin{axis}[
    axis equal image,
    xmin=2.6637, xmax=3.91588,
    ymin=1.11838, ymax=2.37056,
    axis lines=box,
    ylabel={\shortstack{\LARGE \scenarioTag{1}\\y [m]}},
    colorbar,
    colormap name=inferno,
    point meta min=0.245836,
    point meta max=1.55439,
    enlargelimits=false,
    title={\LARGE\RPS{}}
  ]
    \addplot graphics [includegraphics cmd=\pgfimage, xmin=2.6637, xmax=3.91588, ymin=1.11838, ymax=2.37056] {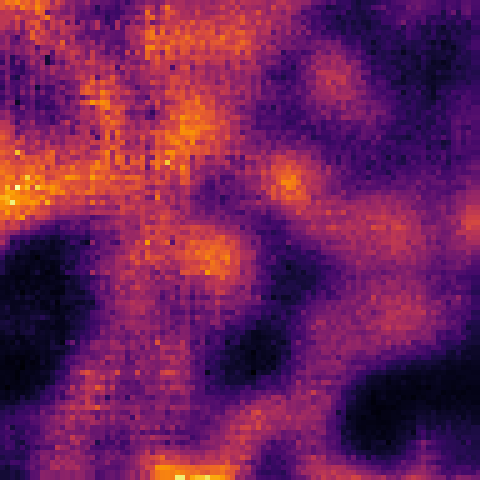};
  \end{axis}
\end{tikzpicture}%
        }
    \end{subfigure}\hfill
    \begin{subfigure}[t]{0.3\linewidth}
        \centering
        \resizebox{!}{\heatmapheight}{%
                \begin{tikzpicture}
  \begin{axis}[
    axis equal image,
    xmin=2.66362, xmax=3.9158,
    ymin=1.11831, ymax=2.37048,
    axis lines=box,
    colorbar,
    colormap name=inferno,
    point meta min=1e-06,
    point meta max=48.7536,
    enlargelimits=false,
    title={\LARGE \GEO{}}
  ]
    \addplot graphics [includegraphics cmd=\pgfimage, xmin=2.66362, xmax=3.9158, ymin=1.11831, ymax=2.37048] {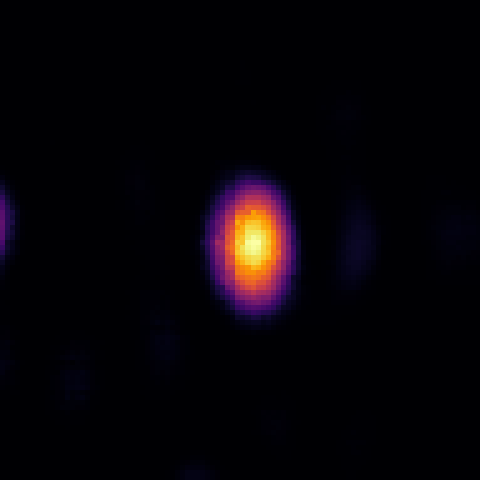};
    \draw[cyan, line width=1pt] (3.27905810546875+0.075, 1.7493585205078126) circle [radius=0.163043/2];
  \end{axis}
\end{tikzpicture}%
        }
    \end{subfigure}\hfill
    \begin{subfigure}[t]{0.3\linewidth}
        \centering
        \resizebox{!}{\heatmapheight}{%
                \begin{tikzpicture}
  \begin{axis}[
    axis equal image,
    xmin=2.6845, xmax=3.93668,
    ymin=1.0804, ymax=2.3717,
    axis lines=box,
    colorbar,
    colormap name=inferno,
    point meta min=1e-06,
    point meta max=58.4388,
    enlargelimits=false,
    title={\LARGE \CSI{}}
  ]
    \addplot graphics [includegraphics cmd=\pgfimage, xmin=2.6845, xmax=3.93668, ymin=1.0804, ymax=2.3717] {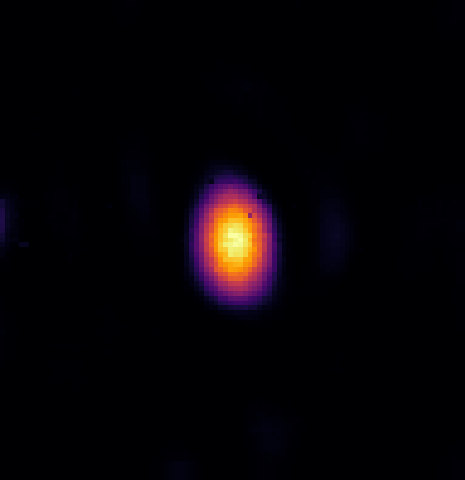};
    \draw[cyan, line width=1pt] (3.32809,1.72864) circle [radius=0.163043/2];
  \end{axis}
\end{tikzpicture}
        }
    \end{subfigure}

    \vspace{0.1cm}
    \begin{subfigure}[t]{0.3\linewidth}
        \centering
        \resizebox{!}{\heatmapheight}{%
            \begin{tikzpicture}
  \begin{axis}[
    axis equal image,
    xmin=2.66375, xmax=3.91592,
    ymin=1.11833, ymax=2.3705,
    axis lines=box,
    xlabel={x [m]},
    ylabel={\shortstack{\LARGE \scenarioTag{2}\\y [m]}},
    colorbar,
    colormap name=inferno,
    point meta min=0.042107,
    point meta max=1.0502,
    enlargelimits=false,
  ]
    \addplot graphics [includegraphics cmd=\pgfimage, xmin=2.66375, xmax=3.91592, ymin=1.11833, ymax=2.3705] {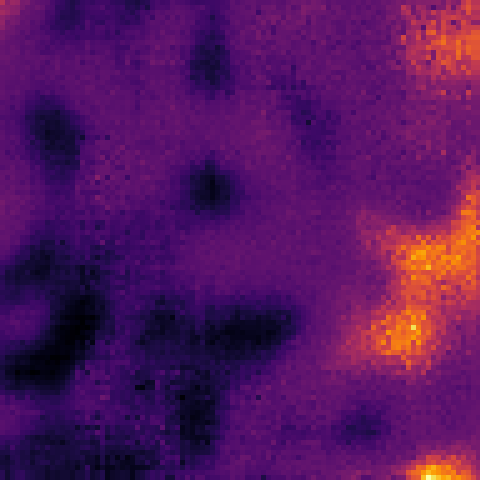};
    \draw[color=cyan, dashed] (3.6,2.4) rectangle (3.0,1.0);
    \draw[fill, cyan] (3.0,2.3705) rectangle (3.6,2.3705-0.02);
    \draw[fill, cyan] (3.0,1.11833) rectangle (3.6,1.11833+0.02);
  \end{axis}
\end{tikzpicture}%
        }
    \end{subfigure}\hfill
   \begin{subfigure}[t]{0.3\linewidth}
        \centering
        \resizebox{!}{\heatmapheight}{%
                \begin{tikzpicture}
  \begin{axis}[
    axis equal image,
    xmin=2.6634, xmax=3.91558,
    ymin=1.11841, ymax=2.37058,
    axis lines=box,
    xlabel={x [m]},
    colorbar,
    colormap name=inferno,
    point meta min=1e-06,
    point meta max=18.4102,
    enlargelimits=false,
  ]
    \addplot graphics [includegraphics cmd=\pgfimage, xmin=2.6634, xmax=3.91558, ymin=1.11841, ymax=2.37058] {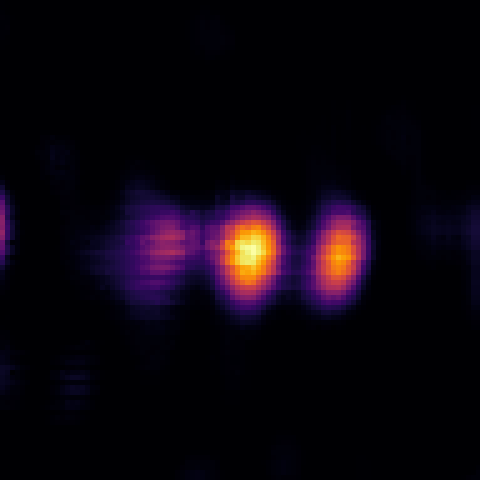};
    \draw[cyan, line width=1pt] (3.25597+0.075,1.77666) circle [radius=0.163043/2];
    \draw[color=cyan, dashed] (3.6,2.4) rectangle (3.0,1.0);
    \draw[fill, cyan] (3.0,2.3705) rectangle (3.6,2.3705-0.02);
    \draw[fill, cyan] (3.0,1.11833) rectangle (3.6,1.11833+0.02);
  \end{axis}
\end{tikzpicture}%
        }
    \end{subfigure}\hfill
    \begin{subfigure}[t]{0.3\linewidth}
        \centering
        \resizebox{!}{\heatmapheight}{%
                \begin{tikzpicture}
  \begin{axis}[
    axis equal image,
    xmin=2.66345, xmax=3.90258,
    ymin=1.13348, ymax=2.37261,
    axis lines=box,
    xlabel={x [m]},
    colorbar,
    colormap name=inferno,
    point meta min=1e-06,
    point meta max=34.1312,
    enlargelimits=false,
  ]
    \addplot graphics [includegraphics cmd=\pgfimage, xmin=2.66345, xmax=3.90258, ymin=1.13348, ymax=2.37261] {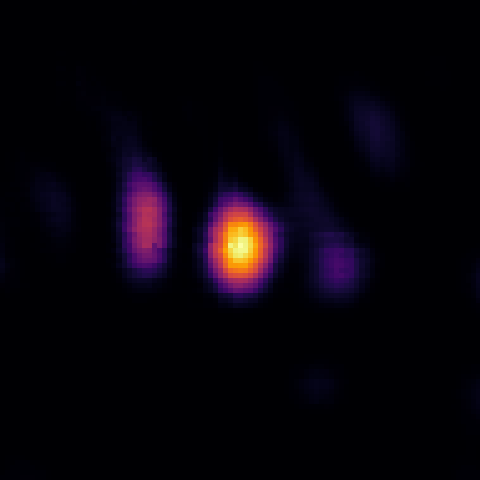};
    \draw[cyan, line width=1pt] (3.28118,1.75131) circle [radius=0.163043/2];
        \draw[color=cyan, dashed] (3.6,2.4) rectangle (3.0,1.0);
    \draw[fill, cyan] (3.0,2.37261) rectangle (3.6,2.37261-0.02);
    \draw[fill, cyan] (3.0,1.13348) rectangle (3.6,1.13348+0.02);
  \end{axis}
\end{tikzpicture}
        }
    \end{subfigure}

    \caption{Heatmaps of measured received \gls{dc} power in \si{\micro\watt} for the three transmission strategies: \rps{}, \geo{} beamforming, and \csi{}-based beamforming (columns), evaluated under \gls{los} and \gls{olos} propagation conditions (rows). Note that different colorbar scales are used to reflect the large dynamic range between the transmission strategies. The cyan rectangle highlights the location of the absorbers and the cyan circle depict the target location with a radius of \(\lambda/4\).
}\label{fig:heatmaps}
\end{figure*}
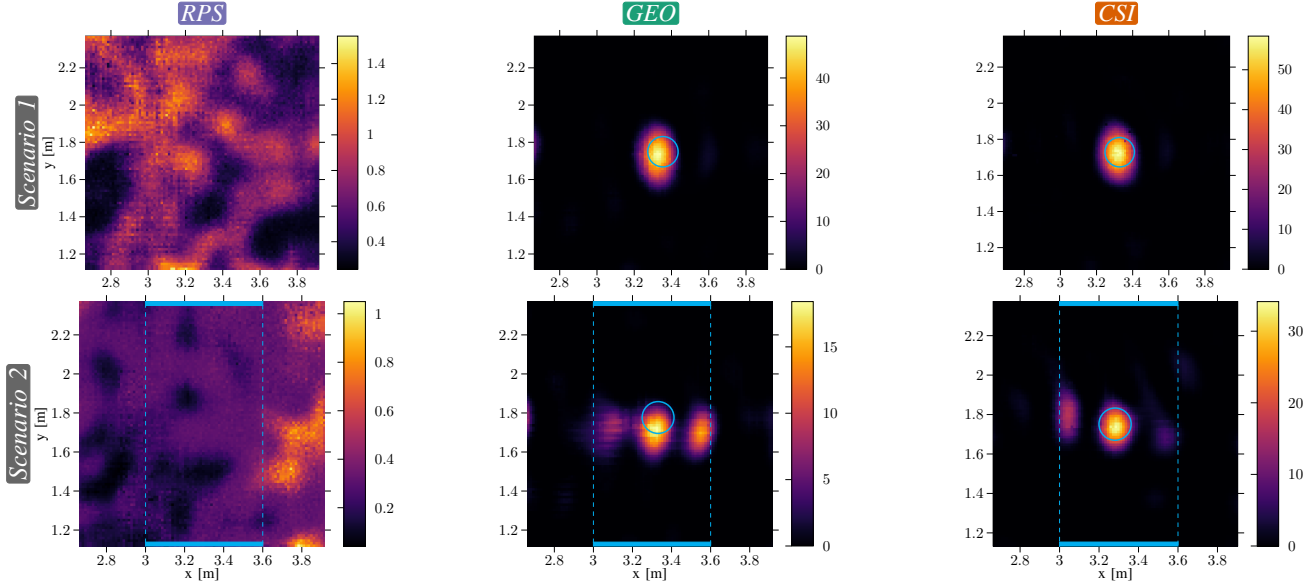

\Cref{fig:heatmaps} shows spatial heatmaps of the measured received \gls{dc} power (in \si{\micro\watt}) for the three transmission strategies: \rps{}, \geo{} beamforming, and \csi{}-based beamforming. Results are shown for both \gls{los} and \gls{olos} propagation conditions. Different colorbar scales are intentionally used to highlight the large dynamic range between non-coherent and coherent transmission strategies.

\scenarioTag{1} \textit{Under \gls{los} conditions}, \RPS{} produces a spatially diffuse power distribution with limited peak power and no pronounced focal point. In contrast, both \GEO{} and \CSI{} beamforming generate a highly localized energy hotspot centered at the target position. The \CSI{}-based approach consistently yields the highest peak power, indicating more accurate phase alignment compared to geometry-only precoding.

\scenarioTag{2} \textit{In the \gls{olos} scenario} with a reflector present, the performance gap between the strategies becomes more pronounced. \GEO{} beamforming partially degrades due to unmodeled reflections and phase errors, resulting in multiple secondary lobes and reduced peak power. \CSI{}-based beamforming remains robust, maintaining a compact focal spot and significantly higher harvested power.

\subsection{Beamforming Gain Analysis}

The measured beamforming gains relative to the \rps{} baseline are summarized in \cref{tab:bf_gain}. The theoretical gain relative to the non-coherent baseline is $10\log_{10}(M)$, corresponding to $20\log_{10}(M)$ relative to a single-antenna transmission.

\scenarioTag{1}  \textit{In the \gls{los} scenario}, both coherent strategies exceed the theoretical benchmark. This can be attributed to the non-power constant RF-to-DC conversion characteristics of the energy harvester (\cref{fig:end-efficiency}) and to residual variance caused by an insufficient number of random-phase realizations used to estimate the \RPS{} baseline. \CSI{}-based beamforming achieves the highest gain of \SI{19.57}{\decibel}, outperforming \GEO{} beamforming by \SI{0.82}{\decibel}.

\scenarioTag{2} \textit{In the \gls{olos} scenario} with reflections, \GEO{} beamforming exhibits a clear performance degradation, underscoring the need to explicitly account for reflected propagation paths. In contrast, \CSI{}-based beamforming achieves a gain of \SI{20.53}{\decibel}, providing an additional \SI{3.83}{\decibel} improvement over geometry-based precoding. This confirms that \CSI{}-based beamforming constructively exploits multipath propagation rather than being degraded by it, even though the absolute received power remains lower than in the \gls{los} case.

\begin{table}[t]
    \centering
    \caption{Measured beamforming gain for \geo{} and \csi{}-based beamforming under \gls{los} and \gls{olos} with a reflector conditions relative to \rps{}, with \(\Delta=\) \,\csi{}\(-\)\text{GEO}.}
    \label{tab:bf_gain}
    \begin{tabular}{lcccc}
        \toprule
        \textbf{Scenario} & \textbf{Theoretical} & \GEO{} & \CSI{} & \(\Delta\)\\
        \midrule
        LoS          & \SI{16.13}{\decibel} & \SI{18.75}{\decibel}  & \SI{19.57}{\decibel} &\SI{0.82}{\decibel}\\
        oLoS + refl.       & \SI{16.13}{\decibel} & \SI{16.70}{\decibel}  & \SI{20.53}{\decibel}  &\SI{3.83}{\decibel}\\
        \bottomrule
    \end{tabular}
\end{table}

\section{Conclusion and Future Work}

This work experimentally evaluated geometry-based multi-antenna \gls{rf} \gls{wpt} using a distributed ceiling-mounted array of \num{41} synchronized transmit antennas operating at \SI{920}{\mega\hertz}. Three transmission strategies were compared: random-phase sweeping as a non-coherent baseline, geometry-based beamforming relying solely on known transmitter-receiver geometry, and \gls{csi}-based beamforming exploiting channel measurements. Spatial power delivery was characterized through two-dimensional scans over a \SI{1.25}{\meter}~$\times$~\SI{1.25}{\meter} area, with harvested \gls{dc} power measured.

Under \gls{los} conditions, geometry-based beamforming achieves similar performance to the \gls{csi}-based beamforming. The gain difference between geometry-based and \gls{csi}-based beamforming in this scenario is limited to \SI{0.82}{\decibel}, indicating that accurate geometric phase alignment is sufficient when propagation is dominated by a single deterministic path.

In contrast, in the \gls{olos} scenario with reflections, geometry-based beamforming degrades to a performance gap of \SI{3.83}{\decibel}, while \gls{csi}-based beamforming maintains a high gain of \SI{20.53}{\decibel}, clearly demonstrating that \gls{csi}-based precoding constructively exploits multipath propagation whereas geometry-only phase compensation does not.

Overall, these results confirm that geometry-based \gls{wpt} is a viable low-overhead strategy for controlled or well-characterized environments, while \gls{csi}-based approaches are preferable in complex indoor settings. Future work will investigate hybrid beamforming strategies that incorporate dominant geometric reflections or limited channel feedback, aiming to recover part of the \SI{3.83}{\decibel} gap in \gls{olos} scenarios while preserving the scalability and low-complexity advantages of geometry-driven designs.

\printbibliography%

@INPROCEEDINGS{Call2607:Experimental,
AUTHOR="Gilles Callebaut and Jarne {Van Mulders}",
TITLE="Experimental Evaluation of Geometry and {Reciprocity-Based} Beamforming
with Large Arrays for {RF} Wireless Power Transfer",
BOOKTITLE="IEEE Wireless Power Technologies Conference and Expo 2026 (WPTCE 2026)",
ADDRESS="Halifax, Canada",
PAGES=5,
DAYS=6,
MONTH=jul,
YEAR=2026
}

@software{github,
  author       = {Callebaut, Gilles and
                  Van Mulders, Jarne},
  title        = {techtile-by-dramco/geometry-based-wireless-power-
                   transfer: v1.0.0
                  },
  month        = mar,
  year         = 2026,
  publisher    = {Zenodo},
  version      = {v1.0.0},
  doi          = {10.5281/zenodo.18952022},
  url          = {https://doi.org/10.5281/zenodo.18952022},
}

@inproceedings{10765709,
	title        = {{Single Versus Multi-Tone Wireless Power Transfer with Physically Large Arrays}},
	author       = {Van Mulders, Jarne and Deutschmann, Benjamin J. B. and Ottoy, Geoffrey and De Strycker, Lieven and Van der Perre, Liesbet and Wilding, Thomas and Callebaut, Gilles},
	year         = {2024},
	booktitle    = {2024 3rd International Conference on 6G Networking (6GNet)},
	pages        = {31--36},
	doi          = {10.1109/6GNet63182.2024.10765709}
}

@inproceedings{9814679,
	title        = {{Location-based Initial Access for Wireless Power Transfer with Physically Large Arrays}},
	author       = {Deutschmann, Benjamin J. B. and Wilding, Thomas and Larsson, Erik G. and Witrisal, Klaus},
	year         = {2022},
	booktitle    = {2022 IEEE International Conference on Communications Workshops (ICC Workshops)},
	pages        = {127--132},
	doi          = {10.1109/ICCWorkshops53468.2022.9814679}
}

@inproceedings{Call2206:Techtile,
	title        = {{Techtile {--} Open {6G} {R\&D} Testbed for Communication, Positioning, Sensing, {WPT} and Federated Learning}},
	author       = {Gilles Callebaut and Jarne {Van Mulders} and Geoffrey Ottoy and Daan Delabie and Bert Cox and Nobby Stevens and Liesbet {Van der Perre}},
	year         = {2022},
	month        = jun,
	booktitle    = {2022 Joint European Conference on Networks and Communications \& 6G Summit (EuCNC/6G Summit): Operational \& Experimental Insights (OPE) (2022 EuCNC \& 6G Summit - OPE)},
	address      = {Grenoble, France},
	days         = {5}
}

@article{ClerckxFoundations2022,
	title        = {{Foundations of Wireless Information and Power Transfer: Theory, Prototypes, and Experiments}},
	author       = {Bruno Clerckx and Junghoon Kim and Kae Won Choi and Dong In Kim},
	year         = {2022},
	journal      = {Proceedings of the IEEE},
	volume       = {110},
	number       = {1},
	pages        = {8--44},
	doi          = {10.1109/JPROC.2021.3132369}
}

@ARTICLE{deutschmann2025physically,
  author={{Deutschmann, Benjamin J. B. and Muehlmann, Ulrich and Kaplan, Ahmet and Callebaut, Gilles and Wilding, Thomas and Cox, Bert and Van der Perre, Liesbet and Tufvesson, Fredrik and Larsson, Erik G. and Witrisal, Klaus},
  journal={IEEE Wireless Communications}}, 
  title={Physically Large Apertures for Wireless Power Transfer: Performance and Regulatory Aspects}, 
  year={2026},
  volume={},
  number={},
  pages={1-8},
  doi={10.1109/MWC.2025.3636246}}

@article{EnergyBall2018,
	title        = {{Energy-Ball: Wireless Power Transfer for Batteryless Internet of Things through Distributed Beamforming}},
	author       = {Xiaoran Fan and Han Ding and Sugang Li and Michael Sanzari and Yanyong Zhang and Wade Trappe and Zhu Han and Richard E. Howard},
	year         = {2018},
	journal      = {Proceedings of the ACM on Interactive, Mobile, Wearable and Ubiquitous Technologies},
	volume       = {2},
	number       = {2},
	pages        = {65:1--65:22},
	doi          = {10.1145/3214268}
}

@article{MittalClosedLoop2024,
	title        = {{An Ultralow-Power Closed-Loop Distributed Beamforming Technique for Efficient Wireless Power Transfer}},
	author       = {Ankit Mittal and Ziyue Xu and Kaden Du and S. Shiva Kumar and Aatmesh Shrivastava},
	year         = {2024},
	journal      = {IEEE Internet of Things Journal},
	volume       = {11},
	number       = {19},
	pages        = {31301--31314},
	doi          = {10.1109/JIOT.2024.3416896}
}

@article{ShenDAS2021,
	title        = {{Wireless Power Transfer With Distributed Antennas: System Design, Prototype, and Experiments}},
	author       = {Shanpu Shen and Junghoon Kim and Chaoyun Song and Bruno Clerckx},
	year         = {2021},
	journal      = {IEEE Transactions on Industrial Electronics},
	volume       = {68},
	number       = {11},
	pages        = {10868--10879},
	doi          = {10.1109/TIE.2020.3036238}
}

@article{UnderstandingConcurrentWPT2024,
	title        = {{Understanding Concurrent Radiative Wireless Power Transfer in the IoT: Out of Myth, into Reality}},
	author       = {Ye Liu and Dong Li and Haipeng Dai and Xiaoyuan Ma and Carlo Alberto Boano},
	year         = {2024},
	journal      = {IEEE Wireless Communications},
	volume       = {31},
	number       = {3},
	pages        = {398--405},
	doi          = {10.1109/MWC.022.2200592}
}

@article{YedavalliBlindBeamforming2017,
	title        = {{Far-Field RF Wireless Power Transfer with Blind Adaptive Beamforming for Internet of Things Devices}},
	author       = {Pavan S. Yedavalli and Taneli Riihonen and Xiaodong Wang and Jan M. Rabaey},
	year         = {2017},
	journal      = {IEEE Access},
	volume       = {5},
	pages        = {1743--1756},
	doi          = {10.1109/ACCESS.2017.2666299}
}

@INPROCEEDINGS{11115919,
  author={An, Zhenlin and Shangguan, Longfei and Kaewell, John and Pietraski, Philip and Jamieson, Kyle},
  booktitle={2025 IEEE International Symposium on Dynamic Spectrum Access Networks (DySPAN)}, 
  title={RadioTwin: A Digital Building Material Twin for Wideband, Cross-link, Cross-band Wireless Channel Prediction}, 
  year={2025},
  volume={},
  number={},
  pages={1-10},
  doi={10.1109/DySPAN64764.2025.11115919}}

@ARTICLE{11037283,
  author={Ren, Meixuan and Dai, Haipeng and Zhang, Linglin and Liu, Tang},
  journal={IEEE Transactions on Mobile Computing}, 
  title={Adaptive Charging With Beam Steering}, 
  year={2025},
  volume={24},
  number={10},
  pages={11224-11240},
  doi={10.1109/TMC.2025.3579692}}

@INPROCEEDINGS{10283480,
  author={Deutschmann, Benjamin J. B. and Wilding, Thomas and Graber, Maximilian and Witrisal, Klaus},
  booktitle={2023 IEEE International Conference on Communications Workshops (ICC Workshops)}, 
  title={XL-MIMO Channel Modeling and Prediction for Wireless Power Transfer}, 
  year={2023},
  volume={},
  number={},
  pages={1355-1361},
  doi={10.1109/ICCWorkshops57953.2023.10283480}}

@INPROCEEDINGS{10000769,
  author={Demarchou, Eleni and Psomas, Constantinos and Krikidis, Ioannis},
  booktitle={GLOBECOM 2022 - 2022 IEEE Global Communications Conference}, 
  title={Energy Focusing for Wireless Power Transfer in the Near-Field Region}, 
  year={2022},
  volume={},
  number={},
  pages={4106-4110},
  doi={10.1109/GLOBECOM48099.2022.10000769}}

@ARTICLE{9270598,
  author={Hajimiri, Ali and Abiri, Behrooz and Bohn, Florian and Gal-Katziri, Matan and Manohara, Mohith H.},
  journal={IEEE Journal of Solid-State Circuits}, 
  title={Dynamic Focusing of Large Arrays for Wireless Power Transfer and Beyond}, 
  year={2021},
  volume={56},
  number={7},
  pages={2077-2101},
  doi={10.1109/JSSC.2020.3036895}}

\end{document}